\documentclass[acmtog]{acmart}

\usepackage{graphicx} 
\usepackage{amsmath}
\usepackage{amsfonts}
\usepackage{enumitem}
\usepackage{wrapfig}

\AtBeginDocument{%
  }

\setcopyright{acmlicensed}
\copyrightyear{2025}
\acmDOI{10.1145/3763352}
\acmConference[ACM SIGGRAPH Asia 2025]{}{December 15-18, 2025}{Hong Kong, China}
\acmYear{2025} \acmVolume{44} \acmNumber{6} \acmArticle{} \acmMonth{12}
\begin{document}

\title{Curve-Based Slicer for Multi-Axis DLP 3D Printing}

\author{Chengkai Dai}
\authornotemark[1]
\email{ckdai@cpii.hk}
\affiliation{%
  \institution{Centre for Perceptual and Interactive Intelligence (CPII)
Limited}
  \country{Hong Kong SAR of China}
}

\author{Tao Liu}
\authornote{Joint first authors.}
\email{tao.liu@machester.ac.uk}
\affiliation{%
  \institution{The University of Manchester}
  \city{Manchester}
  \country{United Kingdom}
}

\author{Dezhao Guo}
\email{dzguo@mae.cuhk.edu.hk}
\affiliation{%
  \institution{The Chinese University of Hong Kong}
  \country{Hong Kong SAR of China}
}

\author{Binzhi Sun}
\email{bzsun@cpii.hk}
\affiliation{%
  \institution{The Chinese University of Hong Kong / Centre for Perceptual and Interactive Intelligence (CPII)
Limited}
  \country{Hong Kong SAR of China}
}

\author{Guoxin Fang}
\email{guoxinfang@cuhk.edu.hk}
\affiliation{%
  \institution{The Chinese University of Hong Kong / Centre for Perceptual and Interactive Intelligence (CPII)
Limited}
  \country{Hong Kong SAR of China}
}

\author{Yeung Yam}
\email{yyam@mae.cuhk.edu.hk}
\affiliation{%
  \institution{The Chinese University of Hong Kong / Centre for Perceptual and Interactive Intelligence (CPII)
Limited}
  \country{Hong Kong SAR of China}
}

\author{Charlie C.L. Wang}
\authornote {Corresponding author: charlie.wang@machester.ac.uk (Charlie C.L. Wang).}
\email{charlie.wang@machester.ac.uk}
\affiliation{%
  \institution{The University of Manchester}
  \city{Manchester}
  \country{United Kingdom}
}

\begin{abstract}
This paper introduces a novel curve-based slicing method for generating planar layers with dynamically varying orientations in digital light processing (DLP) 3D printing. Our approach effectively addresses key challenges in DLP printing, such as regions with large overhangs and staircase artifacts, while preserving its intrinsic advantages of high resolution and fast printing speeds. We formulate the slicing problem as an optimization task, in which parametric curves are computed to define both the slicing layers and the model partitioning through their tangent planes. These curves inherently define motion trajectories for the build platform and can be optimized to meet critical manufacturing objectives, including collision-free motion and floating-free deposition. We validate our method through physical experiments on a robotic multi-axis DLP printing setup, demonstrating that the optimized curves can robustly guide smooth, high-quality fabrication of complex geometries.
\end{abstract}

\begin{teaserfigure}
\centering
\includegraphics[width=\textwidth]{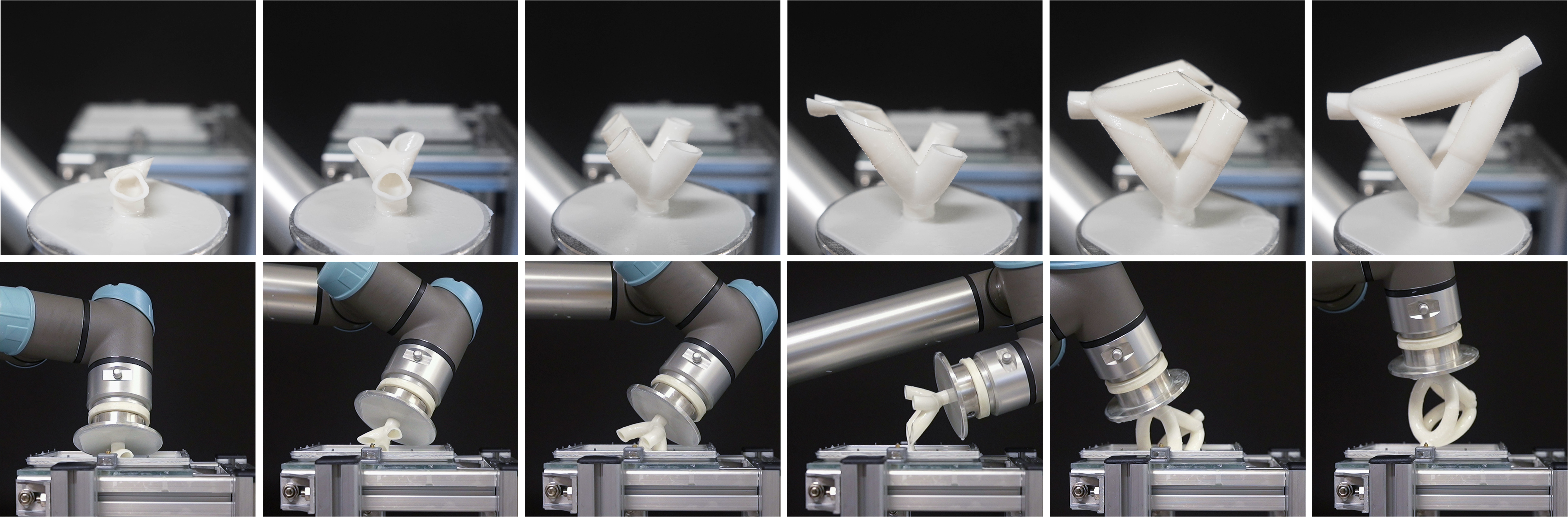}\vspace{-10pt}
\caption{Our method generates curve-based robotic motion trajectories to enable multi-axis DLP 3D printing. The progressive results illustrate the fabrication process of a Toroidal-Tubes model, showcasing the effectiveness of our approach in handling challenging topologies without support structures. 
}\label{fig:teaser}
\end{teaserfigure}



\maketitle

\section{Introduction}

Digital Light Processing (DLP) 3D printing is valued for its high resolution and fast speed, but conventional systems use a single vertical build axis with fixed planar layers. This restriction often requires extensive supports for overhangs, leading to surface damage, increased material waste, and difficult to remove challenges \cite{Zhang2015}. Incorporating multi-axis motion offers a way to overcome these limitations and improve the manufacturability of complex geometries.

Multi-axis strategies have been explored primarily in fused filament fabrication, where dynamically adjusted tool orientations enable support-free printing \cite{Dai2018}, improved surface quality \cite{Etienne2019}, and enhanced mechanical strength \cite{Fang2020,Zhang2022}. Applying this concept to DLP is more difficult: its projection process solidifies an entire resin layer at once, preventing free orientation changes within the layer. Instead, multi-axis DLP adjusts orientations between successive layers (Fig.~\ref{fig:teaser}), exploiting grayscale control of UV light-based curing to modulate local thickness \cite{Luongo2020Microstructure}.

\begin{figure}
\centering
\includegraphics[width=\linewidth]{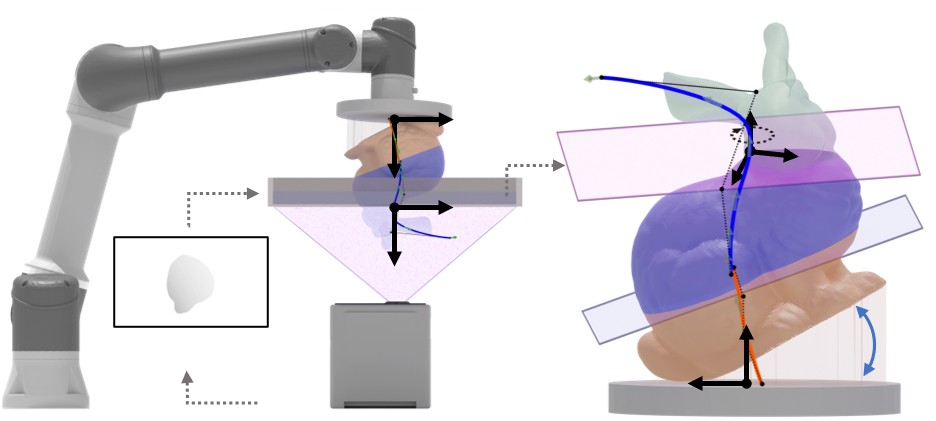}
\put(-129,47){\footnotesize \color{black}$\mathcal{O}_W (\mathcal{O}_T)$}
\put(-129,69){\footnotesize \color{black}$\mathcal{O}_M$}
\put(-187,66){\footnotesize \color{black}Resin Tank}
\put(-216,16){\footnotesize \color{black}Grayscale Image}
\put(-110,70){\footnotesize \color{black}Printing Plane}
\put(-130,15){\footnotesize \color{black}UV-Light}
\put(-130,8){\footnotesize \color{black}Projector}
\put(-78,4){\footnotesize \color{black}$\mathcal{O}_M$}
\put(-42,72){\footnotesize \color{black}$\mathcal{O}_T$}
\put(-90,92){\footnotesize \color{black}$\mathbf{c}_2(t)$}
\put(-48,30){\footnotesize \color{black}$\mathbf{c}_1(t)$}
\put(-15,18){\footnotesize \color{black}$\mathbf{Q}$}
\caption{An illustration for the working principle and the layers of multi-axis DLP 3D printing: the model partition and the multiple trajectories of printing can be formed by multiple curves (displayed in different colors), where material layers are printed at the bottom of resin tank with thickness controlled by the grayscale of the UV-light projection. $\mathcal{O}_M$ represents the frame of MCS and $\mathcal{O}_T$ denotes the frame defined in MCS for each individual printing plane. $\mathcal{O}_W$ is the frame of WCS placed at the center of resin tank for the sake of computation. During printing, $\mathcal{O}_T$ is aligned with $\mathcal{O}_W$ by inverse kinematic computing.
}
\label{fig:ProbDef}
\end{figure}
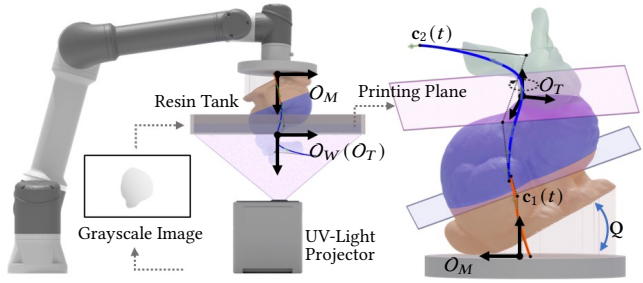

\subsection{Challenges}
The working principle of multi-axis DLP 3D printing can be described as progressively positioning a sequence of oriented planes, $\{ \mathcal{P}_{i=0,1,\ldots,n} \}$, that align with the bottom of the resin tank (as illustrated in Fig.~\ref{fig:ProbDef}). These planes define layers generated by slicing the target model $\mathcal{M}$ along different orientations while several manufacturing objectives must be satisfied to ensure successful fabrication. 
\begin{itemize}
\item \textbf{Collision-Free:} The pose used to realize each planar layer $\mathcal{P}_i$ should ensure the already printed portion of the 3D model remains collision-free with environmental obstacles.

\item \textbf{Connectivity:} Each newly formed layer $\mathcal{P}_i$ must maintain connection with the previously printed structure to prevent floating regions during fabrication. 

\item \textbf{Surface Quality:} To mitigate steep overhangs and minimize staircase artifacts, the angles between the normal vectors of $\mathcal{P}_i$ and the surface of $\mathcal{M}$ should be optimized.

\item \textbf{Smooth Motion:} The planar layers should minimize abrupt orientation changes to ensure smooth, stable, and fast hardware motion. 
\end{itemize}
These objectives must be satisfied for all planar layers. For each planar layer $\mathcal{P}_i$, the already printed part refers to the subset of $\mathcal{M}$ located ‘below’ $\mathcal{P}_i$ in the model coordinate system (MCS) (i.e., the regions highlighted in blue and brown in Fig.~\ref{fig:ProbDef}). 

The primary challenge is to determine feasible layers that are both collision-free and connected, while also optimizing other manufacturing objectives. Although most objectives can be formulated based on layers defined in MCS, ensuring global collision avoidance between the printed model and environmental obstacles requires additional considerations. Specifically, this involves the local frame defined on each layer and the model's setup orientation within the MCS. The setup orientation of a model $\mathcal{M}$ must also be optimized, which can be represented by a transformation $\mathbf{Q} \in \mathbf{SE}(3)$ applied to $\mathcal{M}$ in the MCS. The motion of the build platform is controlled by aligning the local frames of individual layers with the resin tank's frame positioned at the center of the tank, defined in the world coordinate system (WCS). To address these challenges, we propose a curve-based method that optimizes the planar layers and their corresponding local frames simultaneously. 

\subsection{Our Method and Contribution}\label{subsec:OutMethod}
The slicing problem of multi-axis DLP 3D printing involves computing an optimized sequence of planar layers and their corresponding frames. This task can be viewed as a trajectory planning problem that must satisfy the aforementioned manufacturing objectives. In this work, we propose to represent the trajectory as a B\'{e}zier curve $\mathbf{c}(t)$ ($\forall t \in [0,1]$), where its tangent planes define the slicing layers and the corresponding frames are given with the help of the Bishop frames of the curve. The control points $\{\mathbf{p}_k=(x_k,y_k,z_k,\theta_k) \in \mathbb{R}^4 \}$ are treated as variables to optimize the printing trajectory. The first three components of each control point define the shape and frames of the curve in MCS, while the fourth component introduces an additional rotation of the frame in the tangent planes. The manufacturing objectives are encoded as loss functions, defined in terms of these control points and the model's setup transformation in MCS. These objectives are then minimized using a stochastic gradient descent (SGD) solver to achieve an optimal slicing strategy. 

To ensure scalability, we extend the curve-based slicing algorithm to support multiple curves 
$\{ \mathbf{c}_{j=1,2,\ldots,m}(t) \}$, allowing the trajectory in multiple segments to be jointly optimized with the partitions. Specifically, each curve 
$\mathbf{c}_j(t)$ is responsible for generating layers and trajectories for the $i$-th region of $\mathcal{M}$ that is \textit{above} the tangent plane at $\mathbf{c}_j(0)$ and
\textit{below} the tangent plane of the next region at $\mathbf{c}_{j+1}(0)$. Note that the second condition is omitted for the last curve (i.e., when $j=m$). This partitioning divides the model $\mathcal{M}$ into convex regions using the half-spaces defined by the tangent planes at the starting points of the curves (see Fig.~\ref{fig:ProbDef} for an illustration). Similar space partitioning strategies were employed in \cite{Wu2020} for multi-directional 3D printing. However, unlike prior methods where the printing direction remains constant within each partition, our approach optimizes the trajectory within each region. The control points of all curves are optimized jointly, allowing the space partitions to evolve dynamically during the optimization process. Moreover, a refinement scheme is introduced to add curves adaptively.

The technical contributions of our approach are as follows. 
\begin{itemize}
\item 
We formulate the slicing problem for multi-axis DLP 3D printing as a trajectory optimization problem, with the slicing trajectories represented as parametric curves to be optimized. Our formulation enables an efficient solution using gradient-based solvers and smooth realization on a robotic system.

\item 
To ensure printability for points inside a solid, we first define a continuous function that encodes their printing state based on the trajectory curve to be optimized, and then impose constraints requiring that already printed points remain outside environmental obstacles.

\item 
Our computational framework supports slicing with multiple curves (trajectories). Tangent planes at the starting points of these curves define a partition of space, enabling the partition to be jointly optimized together with individual trajectories.

\item 
Manufacturing objectives, such as floating-free, cliff-angle, surface quality, and completeness, can be seamlessly formulated and integrated into our slicer framework for multi-axis DLP 3D printing. 
\end{itemize}
To the best of our knowledge, this is the first slicer that optimizes the continuous change of layer orientations to ensure the manufacturability of complex freeform models in multi-axis DLP 3D printing. We have demonstrated the effectiveness of our approach through computational and physical experiments across a variety of models.

\section{Literature Review}

\subsection{DLP 3D Printing}
DLP 3D printing is a projection-based vat photopolymerization technique that delivers exceptional resolution and accuracy by selectively curing photopolymer resin through controlled, layer-by-layer light exposure \cite{Chaudhary2023DLPSurvey}. Recent advances have significantly expanded the range of photopolymerizable materials suitable for DLP printing \cite{Nam2024}. In addition, sophisticated material processing techniques have been introduced to enable precise control over layer thickness \cite{Luongo2020Microstructure}, colors \cite{Peng2022}, and even spatially varying mechanical properties such as elasticity \cite{kuang2019,Yue2023}. These capabilities are achieved by modulating the grayscale intensity of projected images during resin curing, offering fine-grained control over the printed structure’s material distribution.

\begin{figure}[t]
\centering
\includegraphics[width=\linewidth]{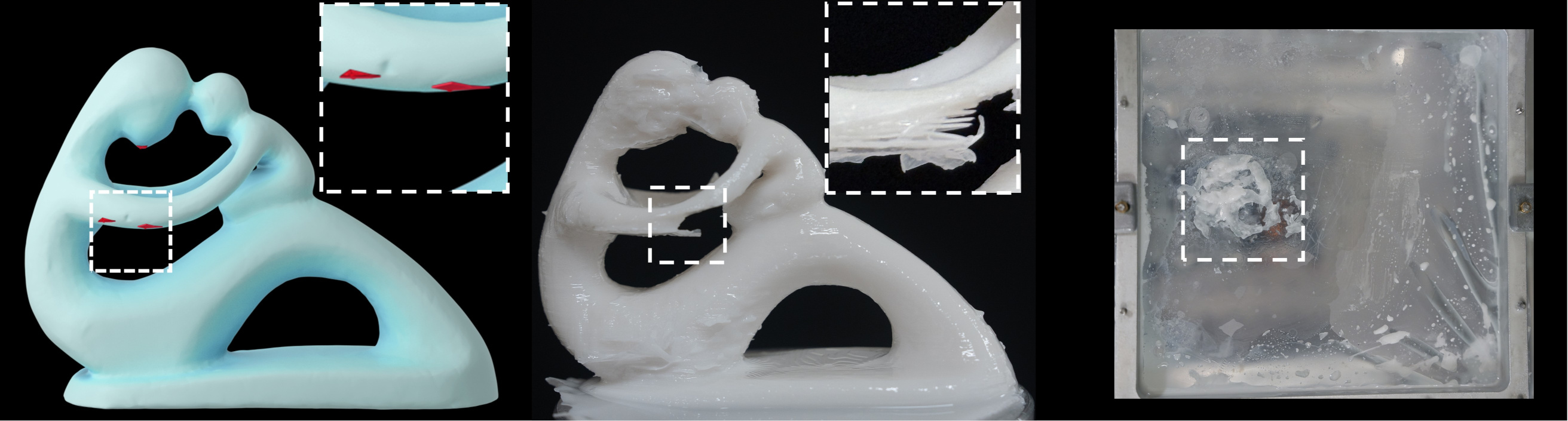}
\put(-242,5){\footnotesize \color{white}(a)}
\put(-161,5){\footnotesize \color{white}(b)}
\put(-80,5){\footnotesize \color{white}(c)}
\caption{Failure case of PPL-based DLP printing due to floating points and regions with large overhangs, highlighted in red color in (a). Directly printing these unsupported regions results in (b) print failure and (c) residual material accumulation at the bottom of the resin tank.
}\label{fig:fertility_failed}
\end{figure}

Despite the advantages of DLP 3D printing in speed and quality, it shares with other parallel planar-layer (PPL) techniques the limitation of requiring support structures for large overhangs and floating points, as illustrated by the failed print in Fig.~\ref{fig:fertility_failed}. This issue is especially problematic for interior regions, such as the tubular features in the Toroidal-Tubes model (Fig.~\ref{fig:teaser}), where supports cannot be removed post-fabrication. Prior strategies include generating slimmed-down supports \cite{Vanek2014CleverSupport, huang2013algorithms,Cao31122025}, optimizing model orientation to reduce visible artifacts \cite{Zhang2015, Liu21MemoryEfficient}, and deforming input geometry to minimize support demand \cite{HU20151Support}. Another drawback of PPL-based printing is staircase artifacts, commonly mitigated by adaptive layer thickness \cite{MAO2019Adaptive}, also explored in filament fusion printing \cite{Wang2015CGF,zhongxu2023}. Decomposing models into sub-components printed along different fixed directions has also been attempted \cite{CAO202160,Wang2016CGF,ZhongZhao23}. However, none of these methods tackles the fundamental cause of both problems -- the globally fixed build direction of planar layers -- which motivates the approach proposed in this paper.

Recently, multiple degrees-of-freedom (DoF) motion in DLP 3D printing has been demonstrated by a robotic system in~\cite{Jigang2021}, which was further extended with a dynamic slicing algorithm based on skeleton-guided strategies~\cite{Jigang2022}. However, their method follows a deterministic approach to handle simple geometries, and the slicing layers are not optimized to improve either manufacturability or surface quality. Separately, various robotic systems have been explored to increase the printable volume in DLP setups~\cite{Yi2016RoboDLP, Yi2018Delta}. Zhao et al.~\cite{Zhao2013Repair} employed DLP printing for adaptive repair of damaged parts through multi-axis tool motion; however, their approach was based on line-wise consolidation rather than layer-wise fabrication. In contrast, we propose a novel computational pipeline that generates planar layers with dynamically varying orientations to tackle the core challenges in DLP-based 3D printing.

\subsection{Multi-Axis 3D Printing}
Multi-axis 3D printing has emerged as a rapidly expanding research area due to the advantages offered by high DoF in motion, which enable material deposition along wireframes of lightweight structures~\cite{Huang2016,Wu2016SIG}, curved layers within solid volumes~\cite{Dai2018}, and shell structures~\cite{Mitropoulou2020}. In general, multi-axis 3D printing can more effectively handle geometries with large overhangs and also enables material deposition along principal stress directions, thereby enhancing mechanical performance~\cite{Fang2020, Tam2017}. These capabilities are often achieved by computing scalar fields whose isosurfaces satisfy both design and manufacturability requirements, and are then extracted as working surfaces for printing.

Such printing layers have also been generated through optimized spatial deformation, forming layers that are either conformal or perpendicular to the model surface to improve surface quality~\cite{Etienne2019}. Building upon this idea, deformation-based optimization has been incorporated into multi-objective frameworks that jointly consider mechanical strength and support-free fabrication~\cite{Zhang2022, Liu2024}. Most recently, this line of research of multi-axis 3D printing has been extended to the simultaneous optimization of structural topology, manufacturable layers, and path orientations for the 3D printing of fiber-reinforced composites~\cite{Liu2025}. However, the layers produced in these works are generally curved surfaces, which are not directly compatible with DLP printing due to its reliance on planar, projection-based fabrication.

On the other aspect, skeleton-based model decomposition was used in~\cite{Wu2016ICRA} to segment an input model so that each segment could be fabricated using PPL-based 3D printing along distinct but constant directions. This approach was later extended to support 3D printing with both 4-axis and 5-axis motion systems~\cite{Wu2020}. However, similar to the previously discussed DLP slicer~\cite{Jigang2022}, these methods do not generate dynamically changing printing directions within sub-regions, nor do they optimize the segmentation based on manufacturing objectives. Notably, the model decomposition and associated layer generation method presented in~\cite{Wu2020} can be interpreted as a special -- yet unoptimized -- case of the solutions produced by our method. In particular, their approach corresponds to the scenario where each sub-region is represented by a straight slicing curve.

\subsection{Robot-Assisted Manufacturing \& Motion Planning}

Robot-assisted manufacturing has seen growing adoption across both additive and subtractive processes. Beyond multi-axis 3D printing enabled by robotic hardware, examples include robotic sculpting~\cite{Ma2020} and curved hot-wire cutting of double-curved surfaces~\cite{Duenser2020}. A central challenge across these processes is avoiding collisions between moving tools and environmental obstacles, which complicates motion planning. Since explicit collision detection can be computationally expensive~\cite{Huang2016}, surrogate models of the environment have been proposed using implicit representations (e.g.,~\cite{Dai2018, Chen2025}). In DLP-based systems, collision risks are even more complex, as they may occur between the resin tank and the dynamically evolving printed part, directly coupling with the slicing process. To address this, we define a state function that represents the time-varying geometry of the part derived from optimized slicing curves, capturing both layer geometry and printing sequence.

More broadly, collision avoidance relates to swept volume computation (e.g.,~\cite{Kim2003Fast,Dziegielewski2010Conservative}), which models the space occupied by moving objects. Recent work extends this with implicit swept volumes using signed distance functions for continuous collision-free trajectory generation~\cite{Wang2024Implicit}. Finally, while our curve-based slicing optimization is conceptually related to curve-based motion planning (e.g.,~\cite{Marcucci2023,Pan2012}), our focus is distinct in targeting manufacturability constraints and print quality optimization.
\begin{figure*}[t]
\centering
\includegraphics[width=\linewidth]{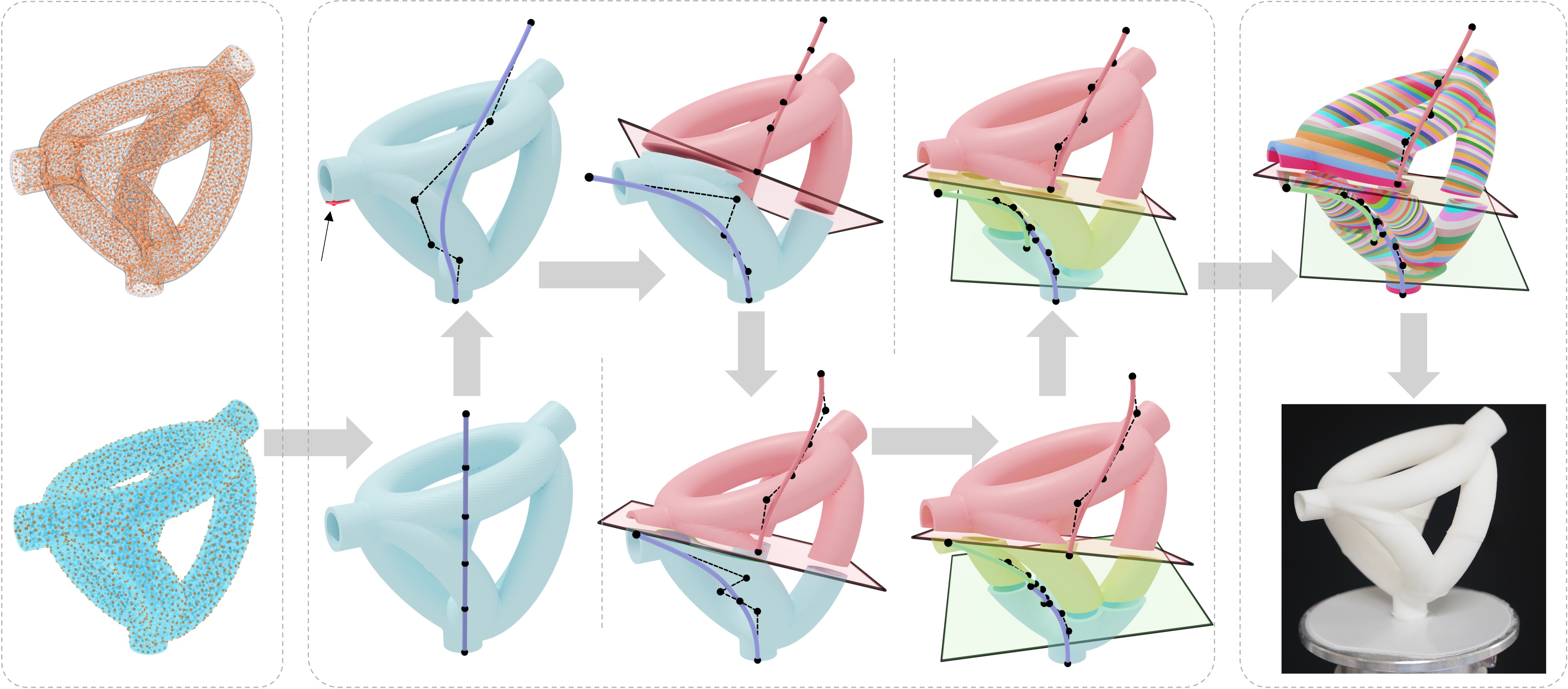}
\put(-385,216){\footnotesize \color{black}\textbf{One Curve}}
\put(-290,216){\footnotesize \color{black}\textbf{Two Curves}}
\put(-192,216){\footnotesize \color{black}\textbf{Three Curves}}
\put(-65,108){\footnotesize \color{black}IK + Printing}
\put(-83,216){\footnotesize \color{black}\textbf{Post-processing}}
\put(-424,90){\footnotesize \color{black}Initialization}
\put(-377,108){\footnotesize \color{black}Optimization}
\put(-284,108){\footnotesize \color{black}Optimization}
\put(-185,108){\footnotesize \color{black}Optimization}
\put(-490,216){\footnotesize \color{black}\textbf{Pre-processing}}
\put(-498,120){\footnotesize \color{black}Solid Sampling: $\Omega \subset \mathcal{M}$}
\put(-505,98){\footnotesize \color{black}Surface Sampling: $\Psi\subset\partial\mathcal{M}$}
\put(-422,135){\footnotesize \color{black}Floating}
\put(-418,128){\footnotesize \color{black}Points}
\put(-380,200){\footnotesize \color{black}$\mathcal{M}_1$}
\put(-378,85){\footnotesize \color{black}$\mathbf{c}_1(t)$}
\put(-338,122){\footnotesize \color{black}Split + Curve Init.}
\put(-247,135){\footnotesize \color{black}$\mathcal{M}_1$}
\put(-285,200){\footnotesize \color{black}$\mathcal{M}_2$}
\put(-262,25){\footnotesize \color{black}$\mathbf{c}_1(t)$}
\put(-265,90){\footnotesize \color{black}$\mathbf{c}_2(t)$}
\put(-228,90){\footnotesize \color{black}Split + Curve Init.}
\put(-194,137){\footnotesize \color{black}$\mathcal{M}_1$}
\put(-145,142){\footnotesize \color{black}$\mathcal{M}_2$}
\put(-190,200){\footnotesize \color{black}$\mathcal{M}_3$}
\put(-158,10){\footnotesize \color{black}$\mathbf{c}_1(t)$}
\put(-190,52){\footnotesize \color{black}$\mathbf{c}_2(t)$}
\put(-140,98){\footnotesize \color{black}$\mathbf{c}_3(t)$}
\put(-116,122){\footnotesize \color{black}Slicing}
\put(-400,8){\footnotesize \color{black}(a)}
\put(-308,8){\footnotesize \color{black}(b)}
\put(-215,8){\footnotesize \color{black}(c)}
\put(-92,116){\footnotesize \color{black}(d)}
\put(-92,8){\footnotesize \color{white}(e)}
\caption{After preparing the computational domain as a set of sample points $\Omega$ for the solid $\mathcal{M}$ and a set of surface sample point $\Psi$, our framework is based on the adaptively applied steps of optimization and splitting. (a) Starting from a straight curve as the trajectory of DLP printing, we optimize the trajectory by changing the positions of its control points (black dots). When critical manufacturing requirement such as floating-free (with floating point highlighted by arrows), the curve is split into two curves to be further optimized (b). Note that the region partition defined by the curves are co-optimized jointly with the curves as trajectories. When hard constraints are still not satisfied 
in the region covered by $\mathbf{c}_1(t)$, $\mathbf{c}_1(t)$ is further adaptively subdivided into two -- i.e., (c) there are three curves in total to be optimized to obtain the slicing result (d). The resultant curves define the trajectories of motion, which can be converted into the execution of robotic system by the computation of inverse kinematics to fabricate the final result (e). 
}\label{fig:algOverview}
\end{figure*}
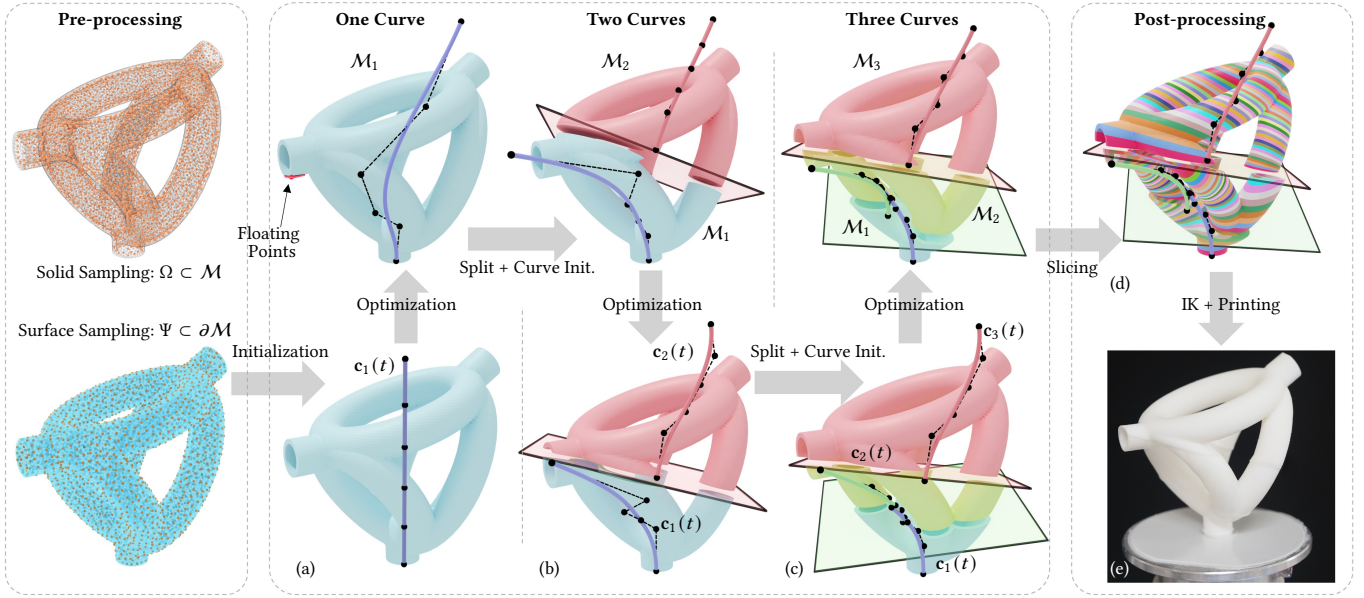

\section{Overview}\label{sec:Overview}
\subsection{Curve Representation, Frames and Variables}\label{subsec:CurveRep}
We represent the motion trajectory of the build platform by a B\'{e}zier curve $\tilde{\mathbf{c}}(t) \in \mathbb{R}^4$ with $t \in [0,1]$ due to its compact formulation and the efficient schemes for evaluation and subdivision by the de casteljau algorithm. Specifically, the curve is given by
$\Tilde{\mathbf{c}}(t) = \sum_{k=0}^n \mathbf{p}_k B_k^n(t)$, 
where the control points ${\mathbf{p}_k = (x_k, y_k, z_k, \theta_k) \in \mathbb{R}^4}$ define both the spatial trajectory and a rotational DoF. The control points are blended together by the Bernstein basis functions $B_k^n(t)$ in order $n$. For clarity, we denote the 3D spatial curve as $\mathbf{c}(t)$ using the $(x,y,z)$-components of $\tilde{\mathbf{c}}(t)$, and the angular profile as $\theta(t)$ using the $\theta$-component. $\mathbf{c}(t) \in \mathbb{R}^3$ is defined in the model coordinate system (MCS).

\begin{wrapfigure}[11]{r}{0.4\linewidth}
\vspace{-10pt}
\centering
\hspace{-25pt}
\includegraphics[width=1.0\linewidth]{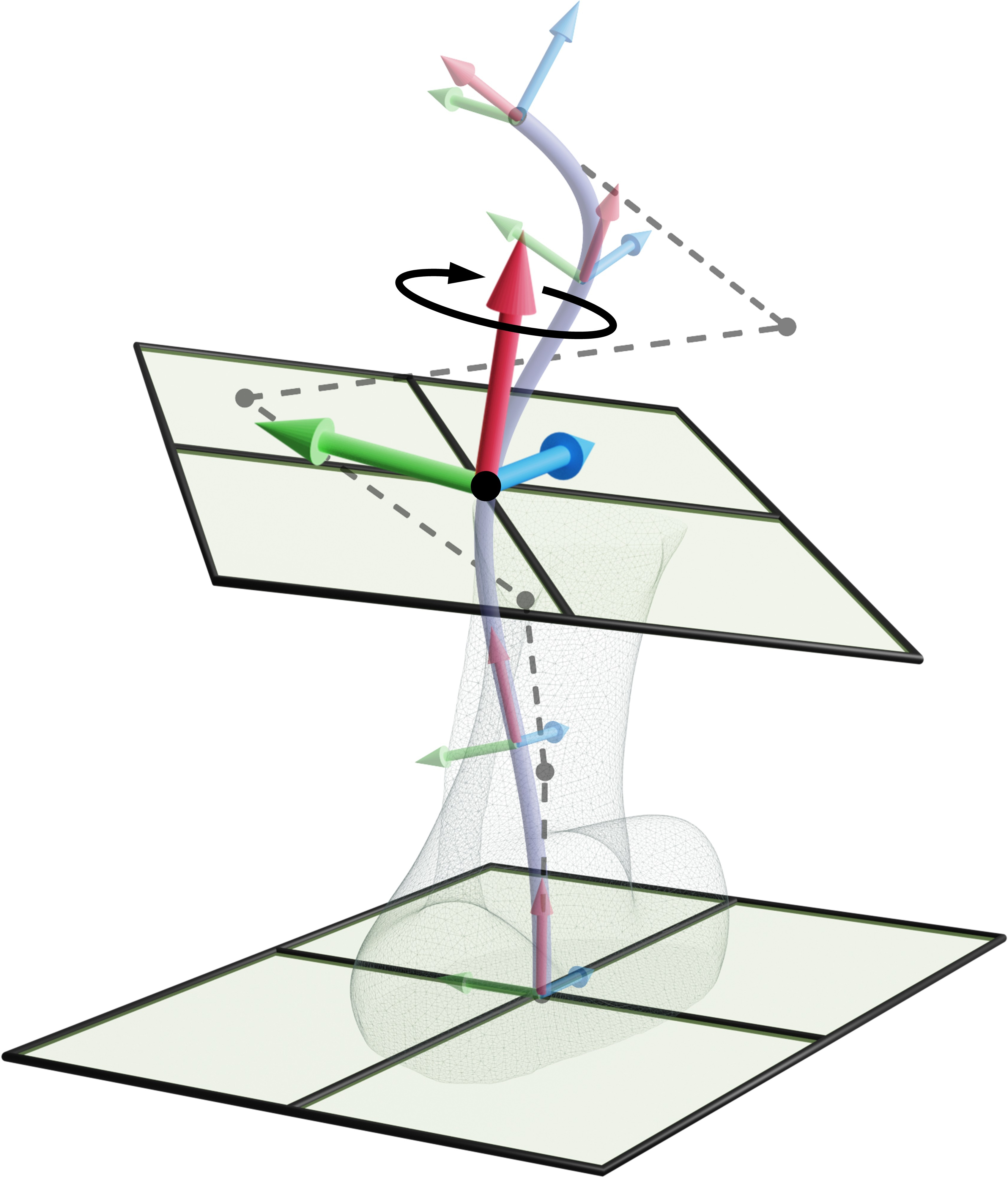}
\put(-57,92){\footnotesize \color{black}$\mathbf{t}(t)$}
\put(-41,67){\footnotesize \color{black}$\mathbf{n}(t)$}
\put(-77,63){\footnotesize \color{black}$\mathbf{b}(t)$}
\put(-73,85){\footnotesize \color{black}$\boldsymbol{\theta}(t)$}
\put(-85,50){\footnotesize \color{black}$\mathcal{P}(t)$}
\put(-47,58){\footnotesize \color{black}$\mathcal{O}_T$}
\put(-50,10){\footnotesize \color{black}$\mathcal{O}_M$}
\end{wrapfigure}
At each parameter $t$, a tangent plane is defined by the point $\mathbf{c}(t)$ and its tangent vector $\mathbf{t}(t) = \dot{\mathbf{c}}(t)$. This plane serves as the dynamically changed printing plane $\mathcal{P}(t)$, whose intersection with the model $\mathcal{M}$ yields a printable layer for DLP fabrication. To define valid local frames along the curve, we adopt the Bishop frame $(\mathbf{t}(t), \mathbf{n}(t), \mathbf{b}(t))$, which avoids issues associated with degeneracies in the normal vector \cite{Wang2008Computation}. We then augment this frame with a rotation of $\theta(t)$ around the tangent vector $\mathbf{t}(t)$, defining the full tooling frame $\mathcal{O}_T$ used for printing (see the insert figure above).

For physical fabrication, we compute a rigid body transformation that aligns $\mathcal{O}_T$ with the world coordinate system (WCS), specifically with a reference frame $\mathcal{O}_W$ placed at the center of the resin tank’s bottom surface as illustrated in Fig.~\ref{fig:ProbDef}. This alignment is represented by a transformation $\mathbf{T}(t) \in \text{SE}(3)$, which maps any point $\mathbf{q} \in \mathbb{R}^3$ (in homogeneous coordinates) in the model $\mathcal{M}$ to a point in WCS as $\mathbf{T}(t) \mathbf{q}$.

Both the printing plane $\mathcal{P}(t)$ and the transformation $\mathbf{T}(t)$ are determined by the control points $\{\mathbf{p}_k\}$, which serve as the design variables in our trajectory optimization for multi-axis DLP printing. Additionally, we can optimize the initial orientation of the model $\mathcal{M}$ by introducing a global transformation $\mathbf{Q} \in \text{SE}(3)$, treated as an additional variable. Under this formulation, a model point $\mathbf{q}$ is transformed to $\mathbf{Q} \mathbf{q}$ in MCS, and further to $\mathbf{T}(t) \mathbf{Q} \mathbf{q}$ in WCS.

\subsection{Optimization for Single and Multiple Curves}
We formulate the curve-based slicing problem as an optimization problem, where the goal is to minimize a set of manufacturing objectives. The optimization variables include the control points $\{\mathbf{p}_k\}$ of the Bézier curve and the global model orientation $\mathbf{Q}$. Formally, it gives
\begin{equation}\label{eq:Optm}
\begin{aligned}
\min_{\{\mathbf{p}_k\},\mathbf{Q}} \quad & \mathcal{L}(\{\mathbf{p}_k\},\mathbf{Q},\mathcal{M}) \\
\text{s.t.} \quad & \Gamma(\{\mathbf{p}_k\},\mathbf{Q},\mathcal{M}) = 0
\end{aligned}    
\end{equation}
where $ \mathcal{L}(\cdot)$ aggregates design goals such as surface quality requirements. Meanwhile, $\Gamma(\cdot)$ ensures the manufacturability of intermediate shapes, including collision-free, connectivity and completeness. The specific formulation about how to evaluate these terms are detailed in Sec.~\ref{sec:ManufacturingObj}.

In practice, it is \textit{not} always able to find a single curve that makes the constraint $\Gamma(\cdot) = 0$ being satisfied. Therefore, we extend the formulation to support multiple curves as
\begin{equation}\label{eq:OptmMultiCurve}
\begin{aligned}
\min_{\{\mathbf{p}^j_k\},\mathbf{Q}} \quad & \sum_{j=1}^m \mathcal{L}_j(\{\mathbf{p}^j_k\},\mathbf{Q},\mathcal{M}) \\
\text{s.t.} \quad & \Gamma_j(\{\mathbf{p}^j_k\},\mathbf{Q},\mathcal{M}) = 0 \quad (j=1,\ldots,m)
\end{aligned}    
\end{equation}
where $\{\mathbf{p}^j_k\}$ are the control points of the $j$-th curve $\mathbf{c}_j(t)$. $\mathcal{L}_j(\cdot)$ and $\Gamma_j(\cdot)$ are piecewise functions that only have non-zero value for the $j$-th region of $\mathcal{M}$ as
\begin{equation}\label{eq:Membership}
\begin{aligned}
    \mathcal{M}_j :=  \{ \mathbf{q} \in \mathcal{M} \, | \,
    & 
    \forall k \leq j,\; (\mathbf{Qq} - \mathbf{c}_k(0)) \cdot \dot{\mathbf{c}}_k(0) \geq 0 \\
    & \land (\mathbf{Qq} - \mathbf{c}_{j+1}(0)) \cdot \dot{\mathbf{c}}_{j+1}(0) < 0 \}.
    \end{aligned} 
\end{equation}
In other words, the tangent planes of curves at their starting points define ordered half-spaces to divide the input model $\mathcal{M}$ in partitions $\{ \mathcal{M}_j \}$. Note that for the last sub-region $j=m$, the last condition will not be applied as the curve $\mathbf{c}_{m+1}(\cdot)$ does not exist. Note that, the transformed position $\mathbf{Qq}$ is employed here for the membership classification.

\subsection{Computational Pipeline}\label{subsec:Alg}
We develop an adaptive refinement algorithm to generate slicing layers via multiple curves. The input model $\mathcal{M}$ is processed as follows:
\begin{itemize}
    \item \textbf{Step 1:} Sample the solid model $\mathcal{M}$ into a set of points $\Omega$ used to evaluate manufacturing objectives such as collision-free motion, connectivity, and completeness.
    
    \item \textbf{Step 2:} Sample $\mathcal{M}$'s surface boundary $\partial \mathcal{M}$ as a point set $\Psi$ for surface quality and support-free objectives.
    
    \item \textbf{Step 3:} Initialize the curve set $\mathcal{C} = \{ \mathbf{c}_1(t) \}$ with a single vertical curve $\mathbf{c}_1(t)$ that spans the height of $\mathcal{M}$.
    
    \item \textbf{Step 4:} Optimize all curves in $\mathcal{C}$ by updating the design variables using a gradient-based solver according to Eq.~\eqref{eq:OptmMultiCurve}.
    
    \item \textbf{Step 5:} For each sub-region $\mathcal{M}_j$, check whether the condition $\Gamma_j(\cdot) = 0$ is satisfied along the trajectory defined by $\mathbf{c}_j(t)$. If not, split the curve into two new curves and update $\mathcal{C}$ by replacing $\mathbf{c}_j(t)$ with the resulting pair. 
    
    \item \textbf{Step 6:} Repeat from Step 4 until all curves satisfy the condition $\Gamma_j(\cdot) = 0$.

    \item \textbf{Step 7:} Adaptively slice the model $\mathcal{M}$ in all sub-regions $\{ \mathcal{M}_j \}$ by the tangent planes of $\mathbf{c}_j(t)$ to generate layers satisfying the manufacturable layer thickness. 
\end{itemize}
Figure~\ref{fig:algOverview} illustrates the progressive refinement of the curve set $\mathcal{C}$ when applied to a geometrically complex model, Toroidal-Tubes.

Unlike prior approaches that partition space deterministically (e.g.,~\cite{Wu2020,Wu2016ICRA}), our method co-optimizes the space partition and slicing curves. During the gradient-based optimization in Step 4, the first control points of all curves $\{ \mathbf{c}_j(t) \}$ with $j > 1$ are allowed to move, enabling dynamic partition adjustment. In contrast, the first control point $\mathbf{p}_0^1$ of the initial curve $\mathbf{c}_1(t)$ is kept on the plane of the build platform with $\dot{\mathbf{c}}_1(0)=(0,0,1)$ to ensure the manufacturability of the first layer. The co-optimization is visualized through the evolving curves and space partitions in Fig.~\ref{fig:algOverview}.
\section{Manufacturing Objectives}\label{sec:ManufacturingObj}
This section presents the formulation of the manufacturing objectives, which must be differentiable with respect to the optimization variables -- namely, the control points $\{ \mathbf{p}_k \}$ of a curve and the setup transformation $\mathbf{Q}$ for adjusting an input model's orientation.

\subsection{States of Point in DLP Printing}
The following formulas define the state of each point $\mathbf{q} \in \mathcal{M}$, including its position in WCS, its \textit{local printing direction} (LPD), and its state about whether already printed.

For a point $\mathbf{q} \in \mathcal{M}$ defined in MCS, its position in WCS at time $t$ is given by $\mathbf{T}(t)\mathbf{Qq}$, where $\mathbf{T}(t)$ is the transformation matrix derived from the Bishop frame $\mathcal{O}_T$, as described in Sec.~\ref{subsec:CurveRep}.

To determine whether a point $\mathbf{q}$ has been printed, we define a state function $D(\mathbf{q}, t)$ that indicates whether \( \mathbf{q} \) contacts the bottom of the DLP printing tank at time $t$. The effective working area\footnote{The effective area is determined by the hardware design of a DLP 3D printer, including the power of UV-light projector, the focus length of lens etc.} at the bottom of resin tank is a rectangle with dimension $[-R_X, R_X] \times [-R_Y, R_Y]$ (see Fig.~\ref{fig:collisionIllustration} for an illustration). Incorporating the factor of effective region, the contact state function $D(\mathbf{q}, t)$ is defined as follows:
\begin{equation}
    D(\mathbf{q}, t) := \delta\left((\mathbf{q} - \mathbf{c}(t)) \cdot \dot{\mathbf{c}}(t)\right) R(\mathbf{q}) 
\end{equation}
where \( \delta(\cdot) \) denotes the Dirac delta function, and $R(\mathbf{q})$ defines the effectiveness according to the working area as
\begin{center}
$R(\mathbf{q}) = H(R_X - | (\mathbf{q} - \mathbf{c}(t)) \cdot \mathbf{n}(t)|) H(R_Y - | (\mathbf{q} - \mathbf{c}(t)) \cdot \mathbf{b}(t)|)$. 
\end{center}
\( H(\cdot) \) is the Heaviside step function, and $\mathbf{n}(t)$ and  $\mathbf{b}(t)$ are unit vectors from the Bishop frame. 
A value of $D(\mathbf{q}, t) = 1$ indicates that \( \mathbf{q} \) is cured into solid by the effective area at the bottom of tank centered at $\mathbf{c}(t)$; otherwise, $D(\mathbf{q}, t) = 0$. Based on this, we define the state function \( s(\mathbf{q}, t) \) to indicate whether \( \mathbf{q} \) has been printed:
\begin{equation}\label{eq:StatueOfPnt}
s(\mathbf{q}, t) := \max_{\tau \in [0, t]} D(\mathbf{q}, \tau),
\end{equation}
where \( s(\mathbf{q}, t) = 1 \) means that the point has been printed at time $t$; otherwise, $s(\mathbf{q}, t) = 0$.

We define a differentiable weight function $w(\mathbf{q}, t) \approx s(\mathbf{q}, t)$ in MCS by applying the following approximations:
\begin{itemize}
    \item Replace the Heaviside function $H(x)$ with the sigmoid function $\sigma_\beta(x) = 1/(1 + \exp({-\beta x}))$;
    \item Approximate the Dirac delta function \( \delta(x) \) with the derivative of the sigmoid function $\dot{\sigma}_\beta(x)$;
    \item Replace the max function \( \max_{\tau \in [0, t]} D(\mathbf{q}, \tau) \) with the log-sum-exp function $\text{LSE}_\alpha(x)$.
\end{itemize}
$\alpha, \beta \to \infty$ are scale coefficients controlling the sharpness of the respective approximations.

Many manufacturing objectives in multi-axis DLP 3D printing are defined based on the LPD at a model point $\mathbf{q} \in \mathcal{M}$. Given the trajectory curve $\mathbf{c}(t)$ for DLP printing, the LPD at $\mathbf{q}$ can be obtained by
\begin{equation}\label{eq:LPD}
    \mathbf{d}(\mathbf{q}) := \int_0^1 \dot{\mathbf{c}}(t) D(\mathbf{q},t) dt \approx \int_0^1  \dot{\mathbf{c}}(t) \dot{\sigma}_\beta\left((\mathbf{q} - \mathbf{c}(t)) \cdot \dot{\mathbf{c}}(t)\right) R_{\delta}(\mathbf{q}) dt  
\end{equation}
with
\begin{center}
    $R_{\delta}(\mathbf{q}):= \sigma_\beta \left(R_X - | (\mathbf{q} - \mathbf{c}(t)) \cdot \mathbf{n}(t)|\right) \sigma_\beta \left(R_Y - | (\mathbf{q} - \mathbf{c}(t)) \cdot \mathbf{b}(t)|\right)$.
\end{center}
The same approximation of $\delta(\cdot)$ and $H(\cdot)$ as above is applied.

\subsection{Collision Avoidance}\label{sec:Collision}
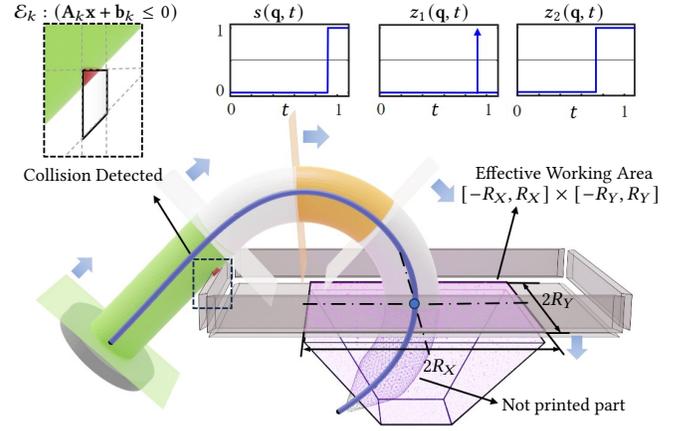
\begin{figure}[t]
\centering
\includegraphics[width=0.45\textwidth]{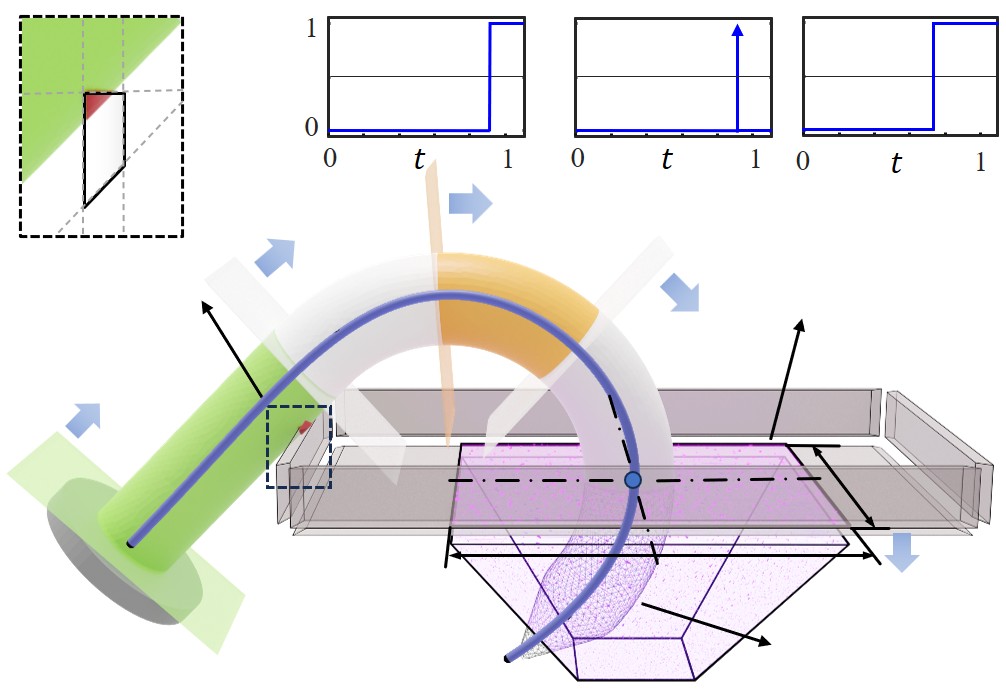}
\put(-237,157){\footnotesize \color{black}$\mathcal{E}_k:(\mathbf{A}_k\mathbf{x}+\mathbf{b}_k\leq 0)$}
\put(-233,95){\footnotesize \color{black}{Collision Detected}}
\put(-36,157){\footnotesize \color{black}$z_2(\mathbf{q},t)$}
\put(-87,157){\footnotesize \color{black}$z_1(\mathbf{q},t)$}
\put(-146,157){\footnotesize \color{black}$s(\mathbf{q},t)$}
\put(-62,96){\footnotesize \color{black} Effective Working Area}
\put(-68,88){\footnotesize \color{black} $[-R_X, R_X]  \times [-R_Y, R_Y]$}
\put(-52,8){\footnotesize \color{black}Not printed part}
\put(-82,23){\footnotesize \color{black}$2R_X$}
\put(-37,48){\footnotesize \color{black}$2R_Y$}
\caption{Illustration of collision detection by representing the environmental obstacles as convex polytopes that can defined by a set of half-spaces, where the region falling inside the convex polytope $\mathcal{E}_k$ has been highlighted in red color. The status whether a point has been printed is indicated by a state function $s(\mathbf{q},t)$ (Eq.\eqref{eq:StatueOfPnt}) defined according to the values of $z_1=\delta\left((\mathbf{q} - \mathbf{c}(t)) \cdot \dot{\mathbf{c}}(t)\right)$ and $z_2=R( \mathbf{q})$ in the range of $[0,t]$ -- see the example shown in the upper-right corner.
}\label{fig:collisionIllustration}
\end{figure}

While optimizing the trajectories for 3D printing, it is essential to prevent collisions between the already printed part and any environmental obstacles. Note that the building plate $\mathcal{B}$ (illustrated as the gray cylinder in Fig.~\ref{fig:collisionIllustration}) needs to be considered as an already printed part for collision check. 

We formulate the environmental obstacles as a finite set of convex polytopes $\mathcal{E}=\{\mathcal{E}_1,\dots,\mathcal{E}_N\}$, and each convex polytope $\mathcal{E}_k \subset\mathbb{R}^3$ is defined by a set of $N_k$ half-spaces (see Fig.~\ref{fig:collisionIllustration}):
\begin{equation}\label{eq:HalfSpaces}
\mathcal{E}_k=\{\mathbf{x}\in\mathbb{R}^3 \, | \, \mathbf{A}_k \mathbf{x}+\mathbf{b}_k \leq \mathbf{0}\},
\end{equation}
where $\mathbf{A}_k$ as a tensor in $\mathbb{R}^{N_k \times 3}$  and $\mathbf{b}_k$ as a vector in $ \mathbb{R}^{N_k}$ define the oriented boundary planes of this convex polytopes. Here, $\mathbf{x}$ is defined in WCS. 
If $\mathbf{x}$ lies strictly inside $\mathcal{E}_k$ (i.e., a collision occurs), all components of the vector as $[\mathbf{A}_k\,\mathbf{x}+\mathbf{b}_k]_i$ ($\forall \; i=1,\ldots,N_k$) should be negative, resulting in
\begin{equation}
    \max (\mathbf{A}_k \mathbf{x}+\mathbf{b}_k) < 0  
\end{equation}
where $\max(\cdot)$ denotes the maximum (non-absolute) component of a vector. Based on this observation, we use the Heaviside function $H(\cdot)$ to indicate whether any already printed point is in collision with the environment while considering all convex polytopes:
\begin{equation}
    \int_{\mathcal{M}} \int_0^1 s(\mathbf{Qq}, t) \sum_k H(-\max (\mathbf{A}_k \mathbf{T}(t)\mathbf{Qq} + \mathbf{b}_k)) dt d \mathbf{q}.
\end{equation}
This integral returns \textit{zero} when no collision occurs throughout the printing process. 

After approximating the maximal function and the Heaviside function with the LSE function and the sigmoid function, we define the collision avoidance loss as
\begin{equation}
\mathcal{L}^{CO} := \sum_{\mathbf{q} \in \Omega} \int_0^1 w(\mathbf{Qq}, t) \sum_k  \sigma_{\beta}( - \text{LSE}_{\alpha} (\mathbf{A}_k \mathbf{T}(t)\mathbf{Qq} + \mathbf{b}_k)) dt.
\end{equation}

When computing the optimization for multiple curves, we can define $\mathcal{L}_j^{CO}$ for $\mathbf{q} \in \Omega_j$ in different regions by using both the transformation $\mathbf{T}(t)$ and the weight function $w(\cdot, t)$ defined on their corresponding curve $\mathbf{c}_j(t)$. Here $\Omega_j$ denote the subset of $\Omega$ as $\Omega_j = \Omega \cap \mathcal{M}_j$.

\subsection{Completeness}\label{subsec:Completeness}
The completeness condition ensures that every point $\mathbf{q} \in \mathcal{M}_j$ within a region $\mathcal{M}_j$ has been effectively swept by the bottom of tank along the trajectory defined by $\mathbf{c}_j(t)$. This condition is satisfied when $s (\mathbf{q}, 1) =1$ for all points at $t=1$. Therefore, we define the completeness loss as
\begin{equation}
    \mathcal{L}^{CP}_j := \sum_{\mathbf{q} \in \Omega_j} (1 - s (\mathbf{q}, 1)) \approx \sum_{\mathbf{q} \in \Omega_j} (1 - w (\mathbf{q}, 1)).
\end{equation}
Again, the weight function $w(\cdot, t)$ is defined on its corresponding curve $\mathbf{c}_j(t)$ in different regions.

\subsection{Support-free: Cliff-angle}\label{subsec:CliffAngle}
One of the primary manufacturing constraints in AM is the requirement for support structures beneath regions with significant overhangs. The necessity for support can be evaluated by examining the angle between the surface normal and the local printing direction at each surface point. Specifically, for a surface point $\mathbf{q} \in \partial \mathcal{M}$ with surface normal $\mathbf{n}_{\mathbf{q}}$, the support-free condition is commonly defined as (ref.~\cite{Zhang2022}):
\begin{equation}
    - \mathbf{n}_{\mathbf{q}} \cdot \mathbf{d}(\mathbf{q}) / \| \mathbf{d}(\mathbf{q}) \|  \leq \sin \eta
\end{equation}
where $\eta$ is the material related coefficient. $\mathbf{d}(\mathbf{q})$ is the LPD at $\mathbf{q}$ (i.e., Eq.\eqref{eq:LPD}), which is determined by the trajectory curve $\mathbf{c}(t)$. 

We define the support-free loss based on cliff-angle (CA) for multiple curves as
\begin{equation}
\mathcal{L}^{CA}_j := \sum_{\mathbf{q} \in \Psi_j}  \rho_{\mathbf{q}} \text{ReLU}\left( - \sin \eta - \mathbf{n}_{\mathbf{q}} \cdot \frac{\mathbf{d}_j(\mathbf{q})}{\| \mathbf{d}_j(\mathbf{q}) \|} \right)
\end{equation}
with $\Psi_j$ being the subset of surface sample points as $\Psi_j = \Psi \cap \mathcal{M}_j$ and $\mathbf{d}_j(\mathbf{q})$ denoting the LPD function as Eq.\eqref{eq:LPD} for the $j$-th curve $\mathbf{c}_j(t)$. The function $\text{ReLU}(\cdot)$ is used to penalize violations where the expression inside is positive. A point sample density, denoted as $\rho_{\mathbf{q}}$, is introduced to account for the non-uniformity of samples. It is defined as the average distance between the point $\mathbf{q}$ and its $k$-nearest neighbors with $k=15$.

\subsection{Support-free: Floating}\label{subsec:SupportFree}
A floating point (also known as point overhang \cite{Vanek2014CleverSupport}) represents an isolated region that lacks supporting structure from below. This occurs when a sample point lies below all of its neighbors along the LPD, forming a local minimum in the geometry. Formally, for a surface sample point $\mathbf{q} \in \partial \mathcal{M}$ with its $k$-nearest neighbors denoted by $\mathcal{N}_{\mathbf{q}}$, it becomes a floating point if 
\begin{equation}
    (\mathbf{p}-\mathbf{q}) \cdot \mathbf{d}(\mathbf{q}) \geq 0   \quad (\forall \mathbf{p} \in \mathcal{N}_{\mathbf{q}}).
\end{equation}
Therefore, the objective of no floating can be reformulated as a requirement to $\mathbf{d}(\mathbf{p})$ as 

\begin{equation}
- \max_{\mathbf{p} \in \mathcal{N}_{\mathbf{q}}} ((\mathbf{q}-\mathbf{p}) \cdot \mathbf{d}(\mathbf{p})) < 0.
\end{equation}
By Eq.\eqref{eq:LPD}, we can then define the floating prevention loss as follows.
\begin{equation}
\mathcal{L}^{FL}_j := \sum_{\mathbf{q} \in \Psi_j} \sigma_{\beta}\left(  - \underset{\mathbf{p} \in \mathcal{N}_{\mathbf{q}}}{\text{LSE}_{\alpha}} \left( (\mathbf{q}-\mathbf{p}) \cdot \mathbf{d}_j(\mathbf{p}) \right)  \right)
\end{equation}
where the $\max (\cdot)$ function is approximated by the LSE function. Again, this is defined on a subset $\Psi_i$ to support the optimization with multiple curves.

\subsection{Surface Quality}\label{subsec:SurfQuality}
Surface quality must be considered to prevent staircase artifacts on the printed model’s surface. As analyzed in \cite{Etienne2019}, maintaining the angle between the surface normal $\mathbf{n}_{\mathbf{q}}$ at a surface point $\mathbf{q} \in \partial \mathcal{M}$ and its LPD $\mathbf{d}(\mathbf{q})$ close to perpendicular is essential.

Using Eq.\eqref{eq:LPD}, a surface quality loss can be defined accordingly for each region as:
\begin{equation}
\mathcal{L}^{SQ}_j := \frac{A_s}{A_{\Psi}} \sum_{\mathbf{q} \in \Psi_j^s} \rho_{\mathbf{q}} \left| \mathbf{n}_{\mathbf{q}} \cdot \frac{\mathbf{d}_j(\mathbf{q})}{\| \mathbf{d}_j(\mathbf{q}) \|} \right|
\end{equation}
The surface quality loss is only applied to user-selected regions $\Upsilon$ in practice as $\Psi_j^s = \Psi_j \cap \Upsilon$, and the loss is weighted by the ratio of the selected region's surface area $A_s$ w.r.t. the input model's surface area $A_{\Psi}$.

\subsection{Total Loss}\label{subsec:TotalLoss}
By incorporating all the loss terms defined above, we formulate the objective term in Eq.~\eqref{eq:OptmMultiCurve} as
\begin{equation}
    \mathcal{L}_j := \omega_{SQ} \mathcal{L}^{SQ}_j + \omega_{CA} \mathcal{L}^{CA}_j 
\end{equation}
and the constraint term as
\begin{equation}
    \Gamma_j := \mathcal{L}^{CO}_j + \mathcal{L}^{CP}_j + \mathcal{L}^{FL}_j. 
\end{equation}
Note that the two support-related losses are handled differently: the floating-prevention loss $\mathcal{L}^{FL}_j$ is treated as a hard constraint, while the cliff-angle loss $\mathcal{L}^{CA}_j$ for support-free is only included in the soft objective. This distinction arises from their practical impact on the printing process. Floating leads to complete print failure, making it a critical constraint. In contrast, as entire layers are solidified simultaneously, surfaces with large cliff angles primarily affect surface quality but do not necessarily cause failure in DLP printing. Nonetheless, such regions must still be carefully managed, particularly where the layer thickness is small.
\section{Implementation Details}
The geometry of the target models is sampled with sufficient density while balancing computational efficiency and accuracy. In our implementation, we set the average distance between sample points slightly smaller than the typical layer thickness (i.e., 0.2~mm) to ensure reliable results. We use B\'{e}zier curves with 6 control points in our implementation. Fewer control points offer insufficient degrees of freedom for optimization, while a larger number results in high-order B\'{e}zier curves that are prone to numerical oscillations. When subdivision is applied, each B\'{e}zier curve is split at $t=0.5$ using de Casteljau's algorithm, ensuring that the original curve is exactly represented by the two newly generated sub-curves.

\subsection{Weighting and Optimization Scheme}
Solving the trajectory optimization problem for multi-axis DLP 3D printing, as defined in Eq.~\eqref{eq:OptmMultiCurve}, is particularly challenging due to the hard constraints of manufacturability defined as $\Gamma_j=0$ ($\forall j=1,\ldots,m$). To balance these strict constraints with other optimization objectives, we adopt the following strategy:
\begin{itemize}
\item We formulate the total loss as a weighted sum: $\sum_j \mathcal{L}_j + \omega_\Gamma \sum_j \Gamma_j$, where $\omega_\Gamma$ is a large penalty weight assigned to the constraint terms. Based on empirical tests taken across all examples in this paper, we set $\omega_\Gamma=100$.

\item We employ the Adam optimizer \cite{kingma2014adam}, a widely used stochastic-gradient based machine learning solver, to minimize the total loss.

\item The optimization begins with a learning rate of 1.0e-3, which is halved every epoch with 1000 iterations. For each epoch, optimization resumes from the best result found in the preceding epoch, allowing for progressive refinement.
\end{itemize}
This hybrid approach -- combining weighted loss formulation with adaptively reduced learning rate -- gives effective optimization of multi-curve trajectories while maintaining strict adherence to manufacturability constraints in our experimental tests.

To enable differentiable optimization, we approximate Heaviside and Dirac delta functions using smoothed variants. The sharpness of these approximations is controlled by tunable parameters. In our implementation, we set $\alpha=\text{1e+3}$ and $\beta=\text{1e+3}$. These values are chosen based on prior literature and validated through empirical tuning. The corresponding weights for different manufacturing objectives are set as $\omega_{SQ}=0.1$ and $\omega_{CA}=1.0$ to balance surface quality and support-free criteria. Finally, the material-specific parameter for cliff-angle evaluation is set to $\eta = 0.698$ (corresponding to a $40^\circ$ cliff-angle) across all experiments. This value is derived from our empirical measurements using physical DLP printing with photopolymer resin.

Our optimization algorithm proceeds iteratively. The iteration terminates when the total loss stops decreasing in the subsequent epoch and all hard constraints are satisfied.

\subsection{Curve Initialization}\label{subsec:CurveInit}
The single-curve optimization is usually initialized using a vertical straight line, with the last control point of $\mathbf{c}(t)$ positioned slightly above the target model. This provides a more favorable starting condition for minimizing the completeness loss. For the $\theta$-components, we simply assign them as zero -- e.g., no additional rotation for initialization.

When splitting a single curve into two (i.e., Step 5 of our algorithm presented in Sec.~\ref{subsec:Alg}), the following heuristics are performed to initialize the curves in preparation for the next round of optimization:
\begin{itemize}
\item \textit{Floating Heuristic}: To address a region with floating points that is covered by the curve $\mathbf{c}_j(t)$, we first move the last control point of $\mathbf{c}_j(t)$ to the closest floating point, $\mathbf{q}_f$. To improve the likelihood of eliminating the floating point, we slightly adjust the control point such that the tangent vector $\dot{\mathbf{c}}_j(1)$ becomes nearly perpendicular to the average direction formed by $\mathbf{q}_f$ and its the $k$-nearest neighbors. Specifically, we require:
\begin{center}
    $\dot{\mathbf{c}}_j(1) \perp \frac{1}{|\mathcal{N}_{\mathbf{q}_f}|} \sum_{\mathbf{p} \in \mathcal{N}_{\mathbf{q}_f}} (\mathbf{p} - \mathbf{q}_f)$,
\end{center} 
where $\mathcal{N}_{\mathbf{q}_f}$ denotes the $k$-nearest neighbors of $\mathbf{q}_f$ in $\mathcal{M}$, and $|\cdot|$ represents the number of elements in the set.  

\item \textit{Margin Heuristic}: For each curve $\mathbf{c}_j(t)$ ($\forall j=2,\ldots,m$), we move the first control point slightly along the inverse tangent vector $-\dot{\mathbf{c}}_j(0)$ by a small distance, creating a marginal overlap between the regions covered by neighboring curves.
\end{itemize}
After applying these initialization heuristics, we observed faster convergence in the optimization process during our numerical experiments. An ablation study for this scheme of curve initialization will be given in Fig.~\ref{fig:tube_withoutCurveInit}.

\subsection{Post-processing}
\begin{wrapfigure}[7]{r}{0.33\linewidth}
\vspace{-1.1\intextsep}
\centering
\hspace{-18pt}
\includegraphics[width=\linewidth]{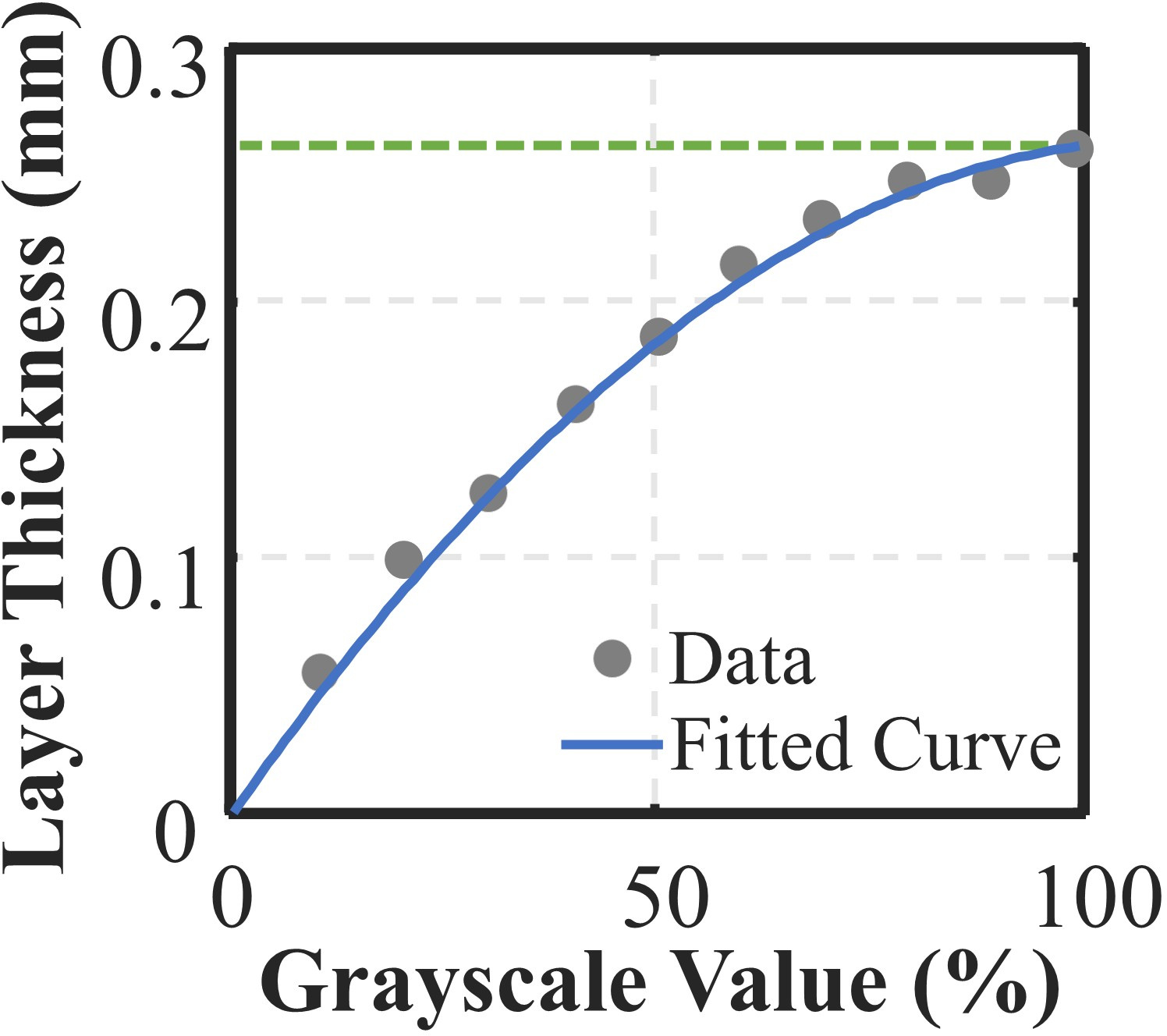}
\put(-62,55){\footnotesize \color{black}$\tau_{\max}$=0.26}
\end{wrapfigure}
The final layers of multi-axis DLP 3D printing are determined by sampling the trajectory curves $\mathbf{c}_j(t)$ densely and slicing the solid model $\mathcal{M}_j$ using the tangent planes of neighboring sample points (e.g., $\mathbf{c}_j(t_c)$ and $\mathbf{c}_j(t_c + \Delta t)$) with a very small step size $\Delta t$, such as 1.0e-3. While the layer thickness during consolidation can be controlled by the grayscale values of the pixels in the projected images -- see the curve shown in the insert figure for a constant exposure time as 5~seconds per layer, the working principle of DLP printing imposes a maximum manufacturable thickness, $\tau_{\max}$. Note that the curve is material-dependent, and can be obtained via calibration. Let $\mathcal{P}_{\text{current}}$ and $\mathcal{P}_{\text{next}}$ denote the planes formed by the tangent plane at $\mathbf{c}_j(t_c)$ and $\mathbf{c}_j(t_c + \Delta t)$ respectively. For every point $\mathbf{q} \in \mathcal{P}_{\text{current}} \cap \mathcal{M}_j$, if the distance from $\mathbf{q}$ to the plane $\mathcal{P}_{\text{next}}$ exceeds $\tau_{\max}$, we iteratively refine the sampling step $\Delta t$ by halving it until the maximum layer thickness constraint is satisfied. 

After determining the slicing result as solid regions on each slice, these regions are converted into images for projection. The motion of the working platform can be computed using inverse kinematics for robotic arms \cite{Dai2020} or parallel mechanisms with tilting tables \cite{Zhang2021}, to align $\mathcal{O}_{T}$ with $\mathcal{O}_{W}$, which is positioned at the center of the resin tank.
\section{Results and Discussion}
We have implemented the method introduced in this paper by Python and the source code of our pipeline will be released upon the acceptance of this paper. The effectiveness of our approach has been verified by both computational and physical experiments. Details are presented below.

\begin{table*}[t]
\caption{Statistics of computational and physical experiments}
\centering\label{tabCompStatistic}
\footnotesize
\begin{tabular}{l | c c c c || c | c || r || r  || r | r | r || r | r | r }
\hline 
& & & & &\multicolumn{2}{c||}{\# Floating Pnt.} & \# Pnt. & Memory$^\dag$ & \multicolumn{3}{c||}{Computing Time (sec.)} & \multicolumn{3}{c}{Fabrication Info.} \\
\cline{6-7} \cline{10-15}
Model  & Fig.  & \#Curves & Objective & Setup Ort. & Before & After & $\Omega + \Psi$ & CPU (MB)  & Optm.$^\ddag$ & Post-proc. & Total & \#Layers & Weight & Time\\
\hline \hline
Toroidal-Tubes & \ref{fig:teaser}, \ref{fig:algOverview} & $3$ & SF & Fixed & 3 & 0 & 13,564 & 1,607 &  $39.3$ & $3.1$ & $42.4$ & $1,495$ & $57$g & $2.9$h\\
Hook$^*$ & \ref{fig:hook_results} & $1$ & SF & Fixed & 0 & 0 & 9,268 & 1,478 & $4.1$ & $4.2$ & $8.3$ & $1,967$ & $52$g & $3.8$h\\
Yoga & \ref{fig:yoga_results} & $1$ & SF & Fixed & 0 & 0 & 9,881 & 1,496 & $10.2$ & $2.6$ & $12.8$ & $1,108$ & $53$g & $2.1$h\\
Woman & \ref{fig:woman_results}  & $2$ & SF + SQ & Fixed & 8 & 0 & 20,820 & 1,825 & $22.3$ & $2.4$ & $24.7$ & $976$ & $122$g & $1.8$h \\
Bunny & \ref{fig:bunny_results}  & $2$ & SF + SQ & Fixed & 0 & 0 & 33,060 & 2,192 &  $45.6$ & $3.1$ & $48.7$ & $1,148$ & $173$g & $2.2$h \\
Fertility & \ref{fig:fertility_results}  & $1$ & SF & Co-Optm. & 6 & 0 & 13,848 & 1,615 &  $10.8$ & $2.4$ & $13.2$ & $961$ & $84$g & $1.8$h \\
\hline
Armadillo & \ref{fig:failureCase_Armadillo}(a) & $4$ & SF & Co-Optm. & 49 & 1 & 41,804 & 2,432  & $62.1$ &  - & -  & - & - & - \\
\hline
\end{tabular}
\begin{flushleft}
\footnotesize
$^*$~Note that there is no floating point when using the medial axis as the initial curve of the hook model; but the computation also converges when using straight line as the initial curve.\\
$^\dag$~The memory used on GPU by the JAX implementation is pre-allocated with 75\% of the graphics memory as 18,428MB on our hardware.\\
$^\ddag$~The optimization time for examples with multiple curves includes the total time of all stages in adaptive subdivision.
\end{flushleft}
\end{table*}

\subsection{Computational Experiments}
All computational experiments are performed on a workstation with an Intel Xeon Sliver 4416+ CPU (20 cores, 3.9 GHz), 512 GB of system memory, and a single NVIDIA RTX 4090 graphics card, operating under Ubuntu 24.04. 
Gradient computations were performed using the JAX framework, and optimization was carried out by the Adam solver in the Optax library.

\begin{figure}[t]
\centering
\includegraphics[width=\linewidth]{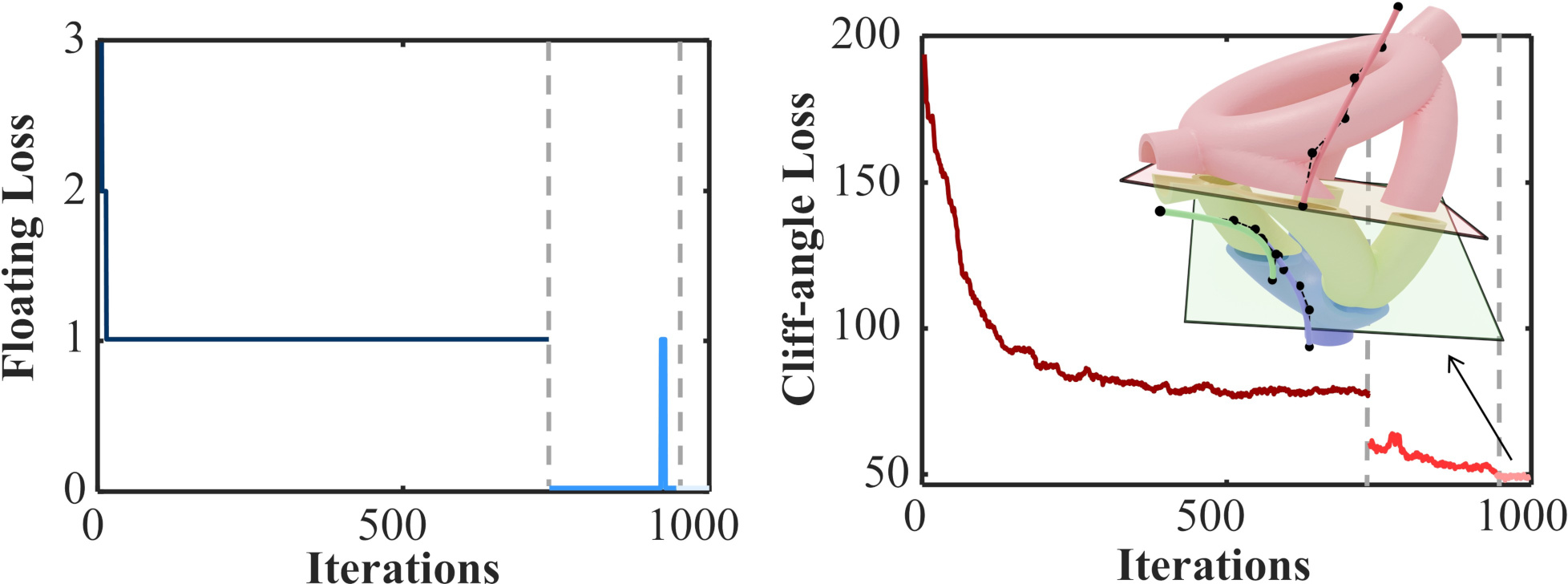}
\put(-244,10){\footnotesize \color{black}(a)}
\put(-120,10){\footnotesize \color{black}(b)}
\caption{Evolution of (a) the floating loss $\mathcal{L}^{FL}$ and (b) the cliff-angle loss $\mathcal{L}^{CA}$ during the optimization process for the Toroidal-Tubes example using 1 to 3 curves. The optimization phases with different numbers of curves are separated by dashed lines.
}\label{fig:curvesLossesTubes}
\end{figure}

\begin{figure}[t]
\centering
\includegraphics[width=0.45\textwidth]{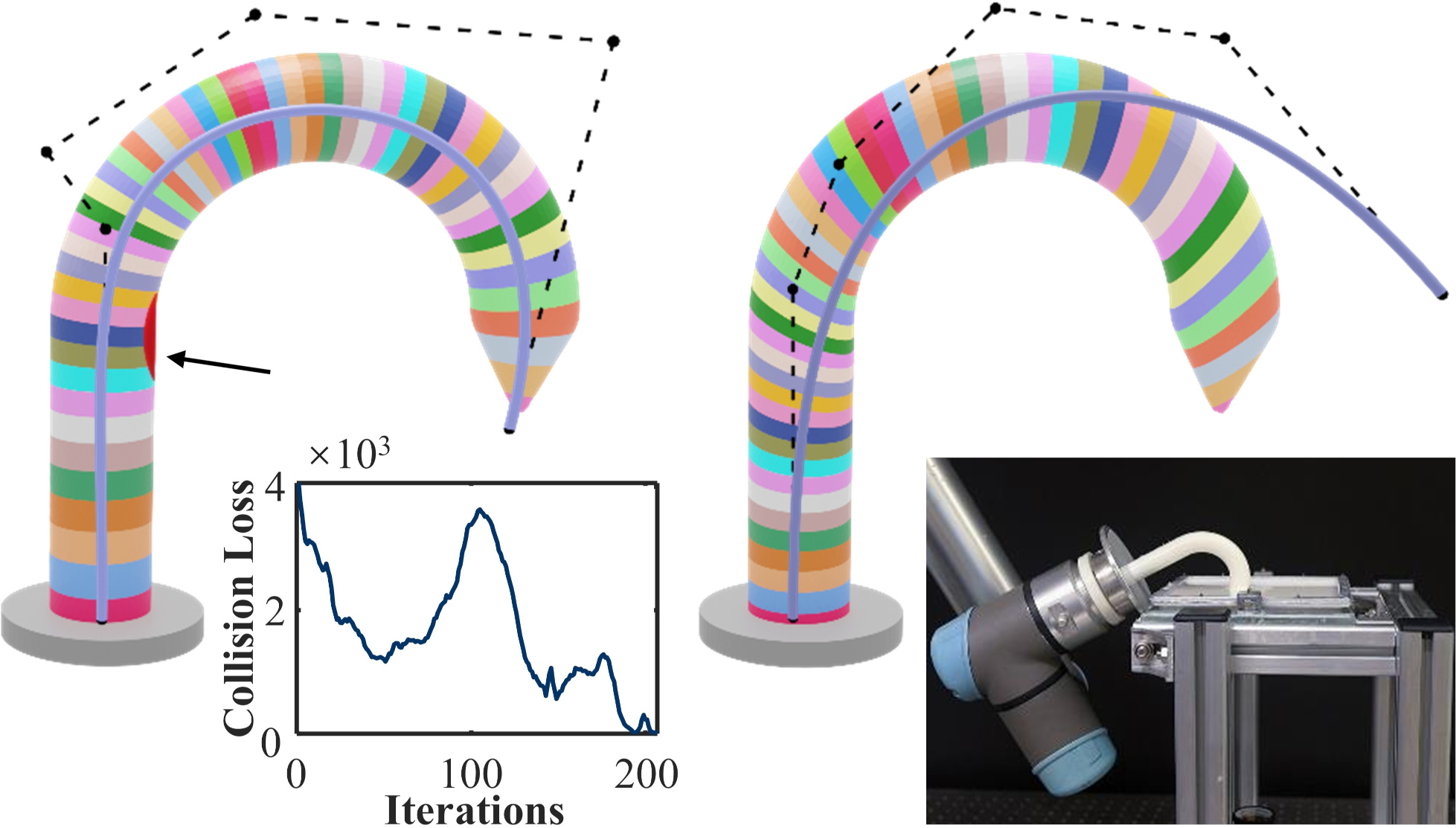}
\put(-232,40){\footnotesize \color{black}(a)}
\put(-123,40){\footnotesize \color{black}(b)}
\put(-205,5){\footnotesize \color{black}(c)}
\put(-100,5){\footnotesize \color{black}(d)}
\put(-186,74){\footnotesize \color{black}Collision}
\put(-186,67){\footnotesize \color{black}Detected}
\caption{The intuitive choice of support-free curve slicing for the Hook model is its medial axis, which however can lead to collision (a) -- see also the collided configuration shown in Fig.~\ref{fig:collisionIllustration}. The issue can be well solved by our optimization with the collision-free curve for slicing and printing as shown in (b). We also provide (c) the convergence curve of the collision loss $\mathcal{L}^{CO}$ and (d) a photo of the collision-free printing process.
}\label{fig:hook_results}
\end{figure}

\begin{figure}[!t]
\centering
\includegraphics[width=\linewidth]{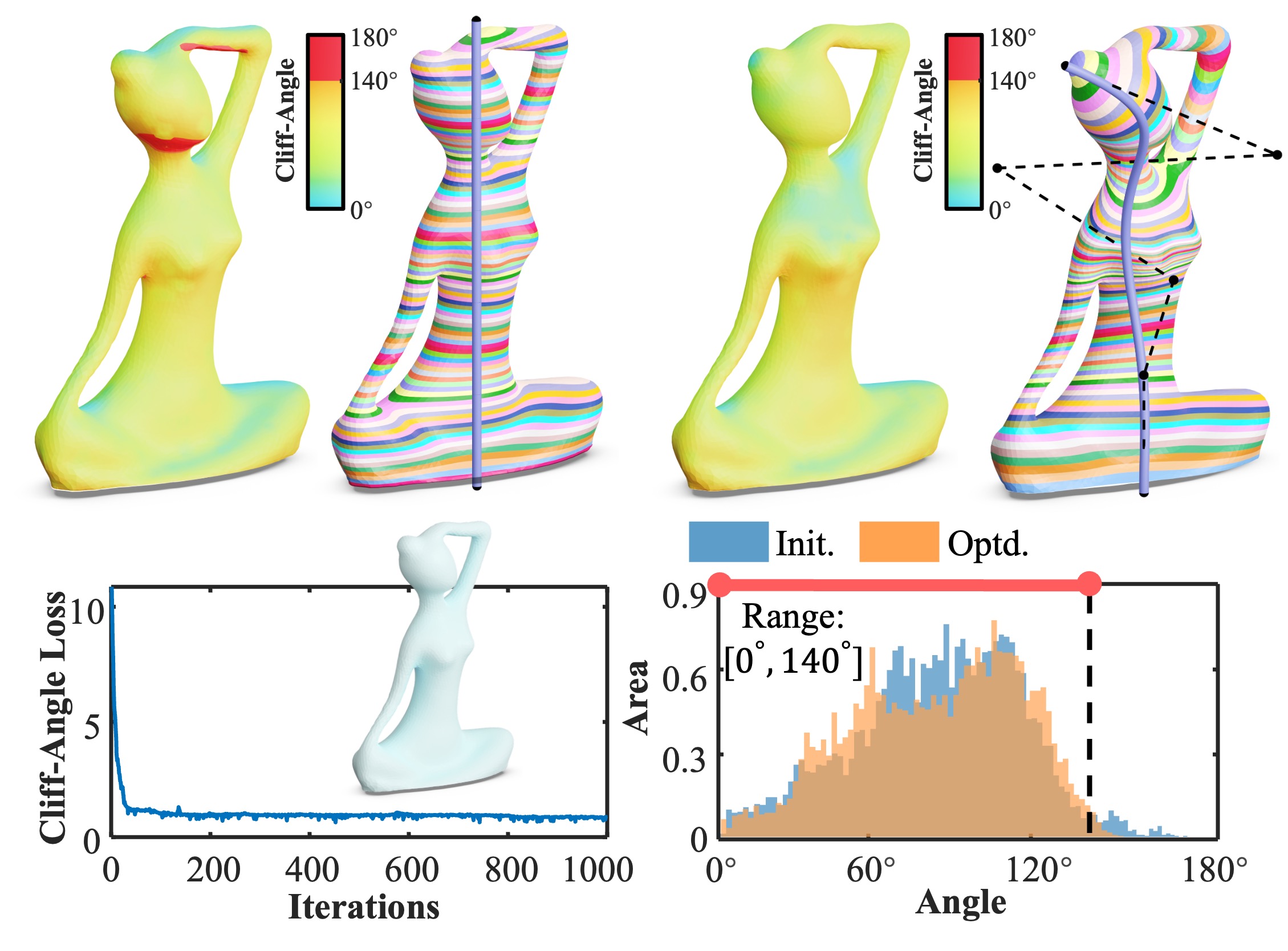}
\put(-243,78){\footnotesize \color{black}(a)}
\put(-123,78){\footnotesize \color{black}(b)}
\put(-243,5){\footnotesize \color{black}(c)}
\put(-123,5){\footnotesize \color{black}(d)}
\put(-15,138){\footnotesize \color{black}$\mathbf{c}(t)$}
\caption{Regions of large overhang in planar DLP printing (highlighted by red color) as shown in (a) can be effectively eliminated by using our curve-based slicer -- see (b) the optimized curve $\mathbf{c}(t)$ and its corresponding control points \& layers, (c) the convergence curve of the cliff-angle loss $\mathcal{L}^{CA}$ and (d) the corresponding histogram of cliff-angles. 
}\label{fig:yoga_results}
\end{figure}

\begin{figure}[t]
\centering
\includegraphics[width=\linewidth]{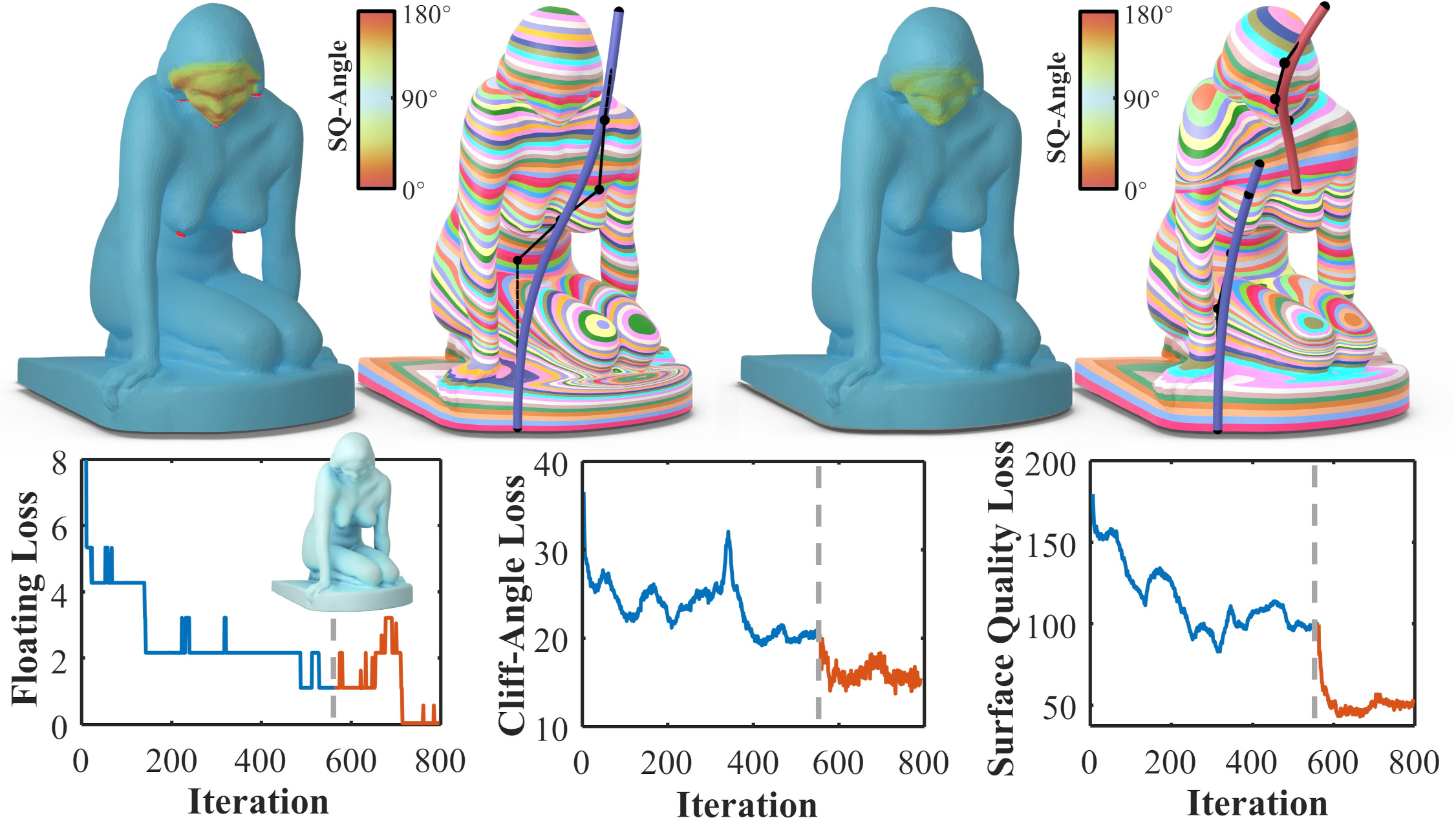}
\put(-241,82){\footnotesize \color{black}(a)}
\put(-120,82){\footnotesize \color{black}(b)}
\put(-241,2){\footnotesize \color{black}(c)}
\caption{Given an input woman model -- the region for surface quality optimization is defined on the face, the results are generated by our method using (a) a single curve -- the floating points exist and highlighted by red dots, and (b) two curves -- no floating point anymore. (c) The convergence curves of computation. The angles between surface normals and the layer orientations have been visualized as color maps on the faces. 
}\label{fig:woman_results}
\end{figure}

\subsubsection{Examples and computing time}
The first example is the Toroidal-Tubes model, which features a complex topology with genus number 4, as shown in Figs.~\ref{fig:teaser} and \ref{fig:algOverview}. In conventional planar layer-based DLP printing, the support structures added inside the tubes cannot be removed after fabrication. For this model, the final DLP layers are generated using three curves — see Fig.~\ref{fig:algOverview} for the curves as trajectories and the corresponding layers. The convergence behaviors of the floating point and the cliff-angle, evaluated with different numbers of trajectories, have been plotted in Fig.~\ref{fig:curvesLossesTubes}. It can be observed that the number of floating points, which is optimized based on the $\mathcal{L}^{FL}$ loss, drops to zero at the end of optimization. The value of $\mathcal{L}^{CA}$ has also been significantly reduced when using three curves. A fixed setup orientation is used throughout this example.

Our method can generate support-free layers using a single curve for a number of models -- for example, the Hook model (Fig.~\ref{fig:hook_results}) and the Yoga model (Fig.~\ref{fig:yoga_results}). In the Hook model example, an intuitive choice of the slicing curve is its medial axis, which however leads to collisions when printing toward the end of the hook (see Fig.~\ref{fig:hook_results}(a)). However, this issue can be effectively resolved by our optimization-based method as shown in Fig.~\ref{fig:hook_results}(b). In the Yoga model example, regions with large overhangs are eliminated after generating layers along a curved trajectory. In these three examples, we neglect the surface quality loss by setting $\omega_{SQ}=0.0$.

\begin{figure}[t]
\centering
\includegraphics[width=\linewidth]{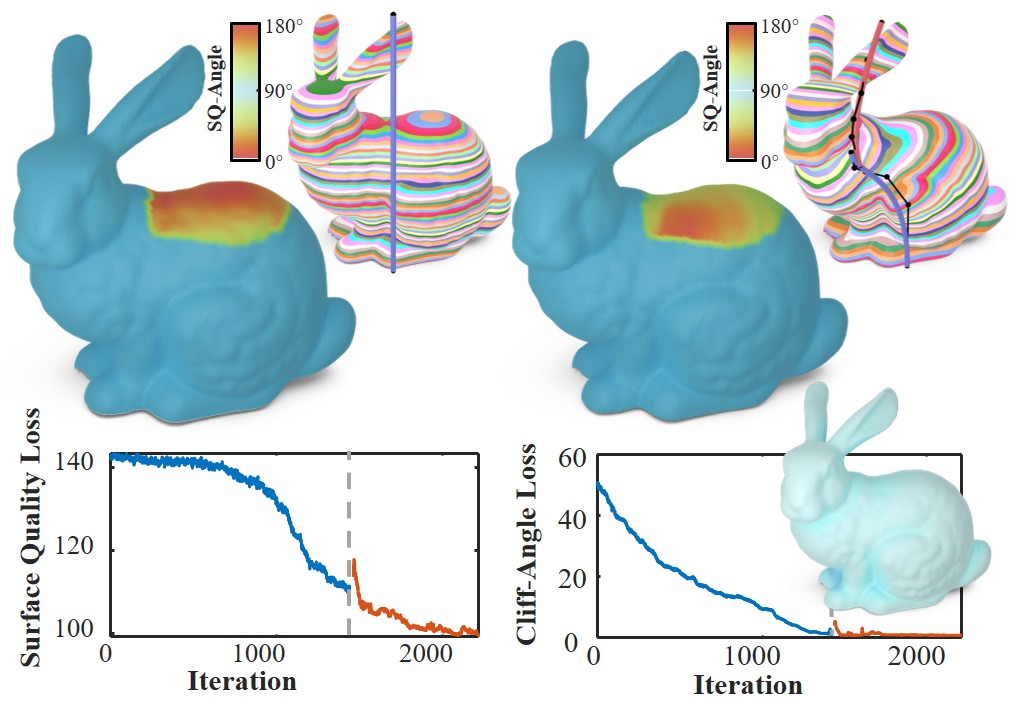}
\put(-243,73){\footnotesize \color{black}(a)}
\put(-120,73){\footnotesize \color{black}(b)}
\put(-243,5){\footnotesize \color{black}(c)}
\caption{The surface quality of the Bunny model (a) in the selected regions on the back can be effectively optimized -- see (b) the new layers and the improved color maps for the distribution of angles between surface normals and the layer orientations. (c) The convergence curves of $\mathcal{L}^{SQ}$ (left) and $\mathcal{L}^{CA}$ (right) are also given.
}\label{fig:bunny_results}
\end{figure}

In the fourth example of a Woman statue model, we incorporate the surface quality objective defined on a user-selected face region (see Fig.~\ref{fig:woman_results}). A manufacturable solution is achieved by subdividing the initial curve into two curves, since a single curve alone cannot fully eliminate floating points. The angles between surface normals and the layers’ normals have been visualized as color maps on the faces. It can be observed that our method can effectively shifts the distribution toward orthogonality. Surface quality optimization is also demonstrated on the Bunny model shown in Fig.~\ref{fig:bunny_results}. Again two curves are employed in this example to realize support-free printing while preserving the surface quality at the back of the Bunny model.

\begin{figure}[t]
\centering
\includegraphics[width=\linewidth]{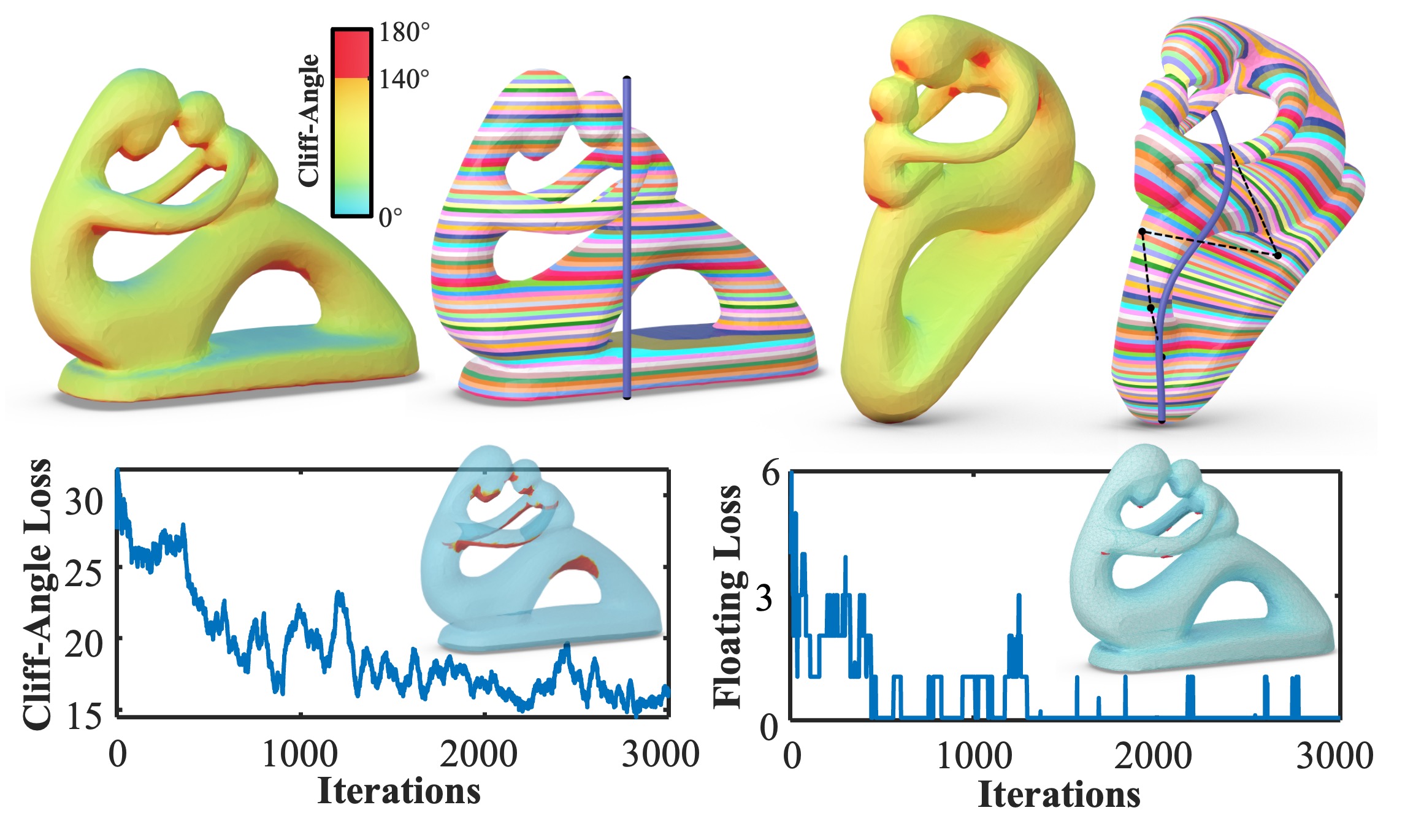}
\put(-245,72){\footnotesize \color{black}(a)}
\put(-103,72){\footnotesize \color{black}(b)}
\put(-245,5){\footnotesize \color{black}(c)}
\caption{There are many challenging regions on the Fertility model with large overhangs -- the regions with large overhangs are highlighted by red color in (a). Our method can optimize the slicing curve and the setup orientation together to reduce most of these regions and eliminate all floating points -- see the resultant curve and its corresponding layers shown in (b). The convergence curves of the cliff-angle loss and the floating loss are given in (c).
}\label{fig:fertility_results}
\end{figure}

The final example is the Fertility model, which poses significant challenges for conventional PPL-based DLP printing. By allowing the setup orientation to be co-optimized, we are able to generate layers for multi-axis DLP printing using a single curve (see Fig.~\ref{fig:fertility_results}). 

The computational statistics for all examples are summarized in Table~\ref{tabCompStatistic}. Overall, our approach demonstrates efficient performance -- all models are completed within one minute. It can be observed that the step of curve optimization takes the majority of computing time while the post-processing step can be completed within few seconds in general.

\begin{figure}[t]
\centering
\includegraphics[width=\linewidth]{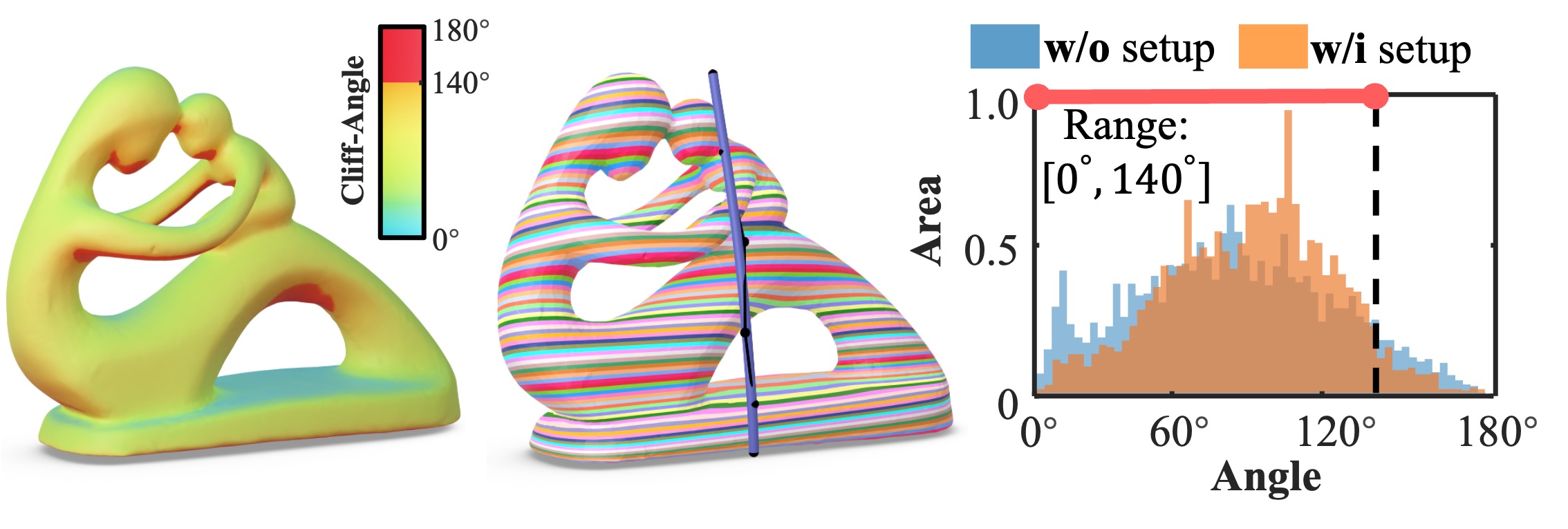}
\put(-245,5){\footnotesize \color{black}(a)}
\put(-90,5){\footnotesize \color{black}(b)}
\caption{Disabling the co-optimization of setup orientation results in large regions with significant overhang, highlighted in red in (a). This is further visualized by (b) the histograms comparing the angles between surface normals and local printing directions. This also leads to floating points.
}\label{fig:fertility_resultsComparison}
\end{figure}

\begin{figure}[t]
\centering
\includegraphics[width=\linewidth]{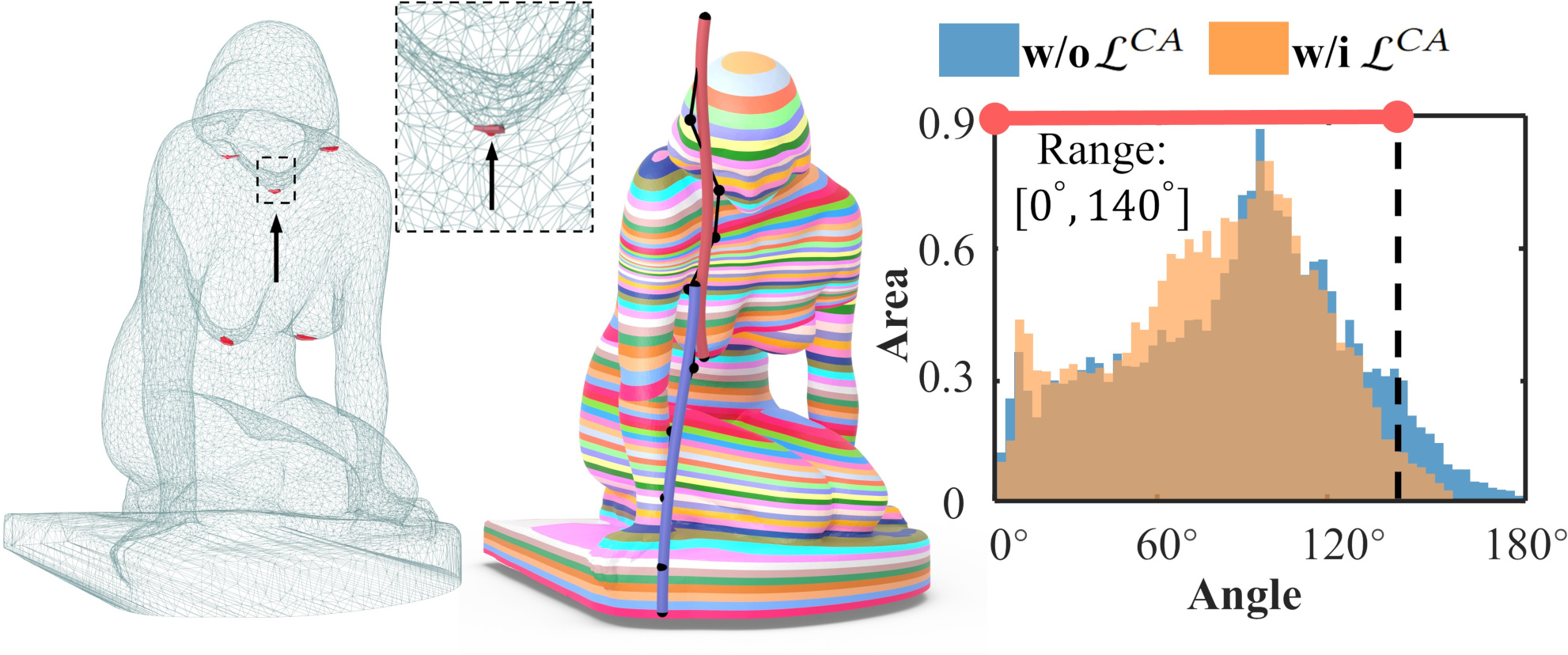}
\put(-243,5){\footnotesize \color{black}(a)}
\put(-95,5){\footnotesize \color{black}(b)}
\caption{Ablation study of two different support-free losses on the Woman model: (a) the result by removing the floating loss $\mathcal{L}^{FL}$ but keeping $\mathcal{L}^{CA}$, and (b) the histograms of cliff-angles with vs. without using the $\mathcal{L}^{CA}$ loss while keeping the floating loss $\mathcal{L}^{FL}$.
}\label{fig:woman_resultsComparison}
\end{figure}

\subsubsection{Ablation study}\label{subsubSec:AblationStudy} 
Co-optimizing the setup orientation $\mathbf{Q}$ along with the curve trajectory for multi-axis DLP printing generally provides the optimizer with a larger solution space, increasing the likelihood of finding a feasible result. For the Fertility model, an ablation study has been taken by turning on / off the co-optimization of setup orientation. From the result shown in Fig.~\ref{fig:fertility_resultsComparison}, we find that it is difficult to effectively reduce the area of overhangs without the co-optimization.  

\begin{figure}[t]
\centering
\includegraphics[width=\linewidth]{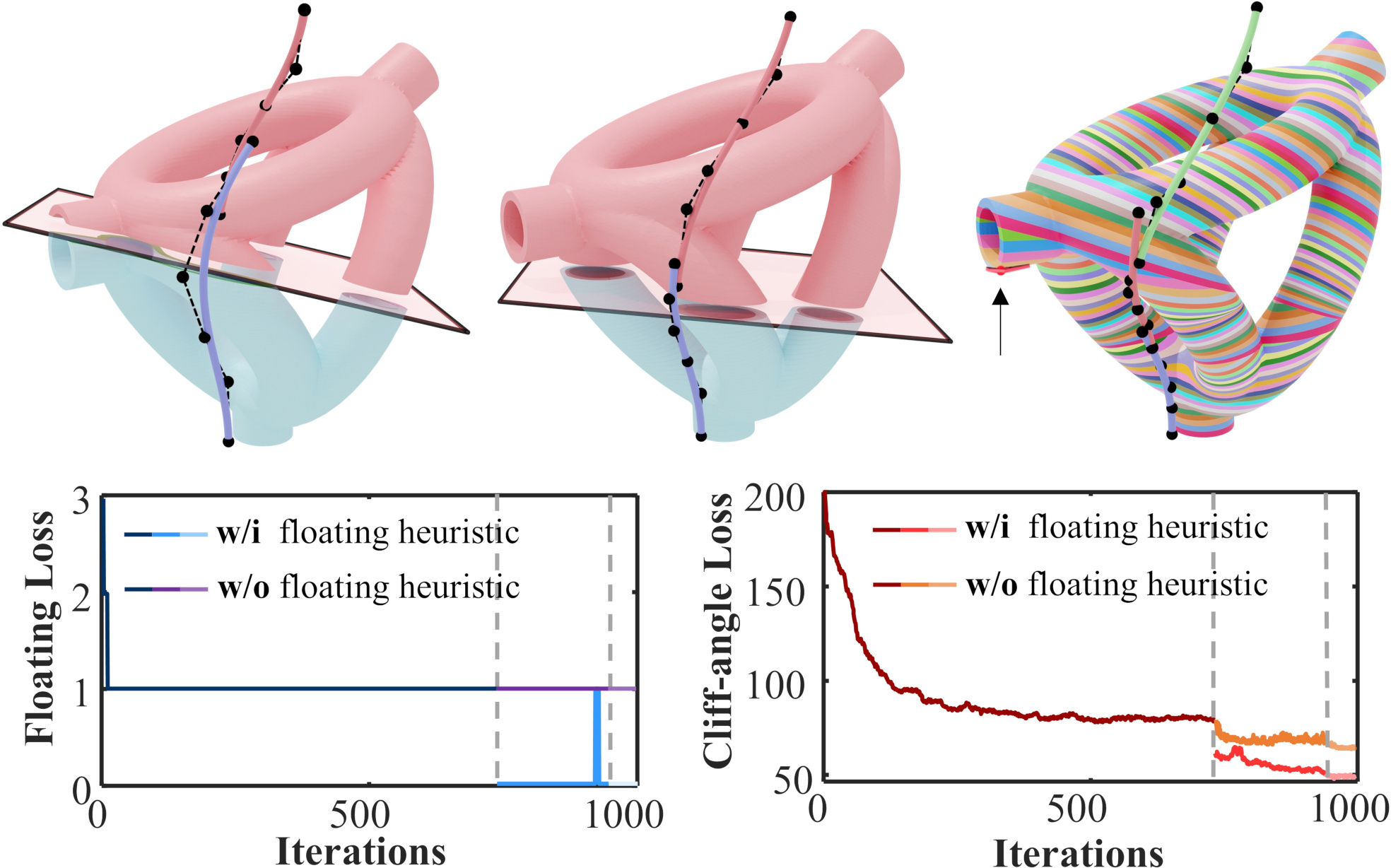}
\put(-240,75){\footnotesize \color{black}(a)}
\put(-155,75){\footnotesize \color{black}(b)}
\put(-75,75){\footnotesize \color{black}(c)}
\put(-88,83){\footnotesize \color{black}Floating Point}
\put(-240,5){\footnotesize \color{black}(d)}
\put(-125,5){\footnotesize \color{black}(e)}
\caption{Ablation study of the floating heuristic employed in the curve initialization scheme: (a) the two curves after initialization without applying the floating heuristic, (b) the optimized curves and the corresponding partitions, and (c) the result of optimization taken on three curves (i.e., floating point cannot be eliminated without the floating heuristic). The computational convergence curves of (d) the floating loss $\mathcal{L}^{FL}$ and (e) the cliff-angle loss $\mathcal{L}^{CA}$ further demonstrate the effectiveness of the floating heuristic.
}\label{fig:tube_withoutCurveInit}
\end{figure}

\begin{figure}[t]
\centering
\includegraphics[width=\linewidth]{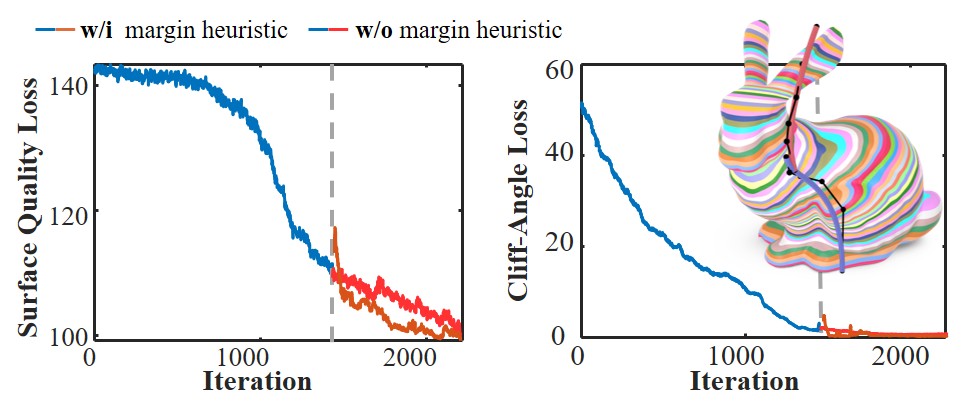}
\put(-243,5){\footnotesize \color{black}(a)}
\put(-120,5){\footnotesize \color{black}(b)}
\caption{Ablation study of the margin heuristic in the curve initialization scheme, demonstrating its effectiveness through the convergence behavior of (a) the surface quality loss $\mathcal{L}^{SQ}$ and (b) the cliff-angle loss $\mathcal{L}^{CA}$ on the Bunny model.
}\label{fig:bunny_result_ablation}
\end{figure}

We further conducted an ablation study to demonstrate the importance of both floating-free losses, $\mathcal{L}^{FL}$ and $\mathcal{L}^{CA}$. On the Woman model, a substantial number of floating points are generated when the $\mathcal{L}^{FL}$ loss is removed (see Fig.~\ref{fig:woman_resultsComparison}(a)). Conversely, when using only $\mathcal{L}^{FL}$ without considering the cliff-angle loss $\mathcal{L}^{CA}$, more regions with large cliff-angles are generated (see the histograms given in Fig.~\ref{fig:woman_resultsComparison}(b)), highlighting the complementary roles of these two objectives.

The other two ablation studies have been taken to verify the effectiveness of our curve initialization scheme. As shown in Fig.~\ref{fig:tube_withoutCurveInit}, when the floating heuristic described in Sec.~\ref{subsec:CurveInit} is not applied, the computation becomes more difficult to converge. 
Similarly, omitting the margin heuristic leads to slower convergence during optimization as shown in Fig.~\ref{fig:bunny_result_ablation}.

\begin{figure}[t]
\centering
\includegraphics[width=\linewidth]{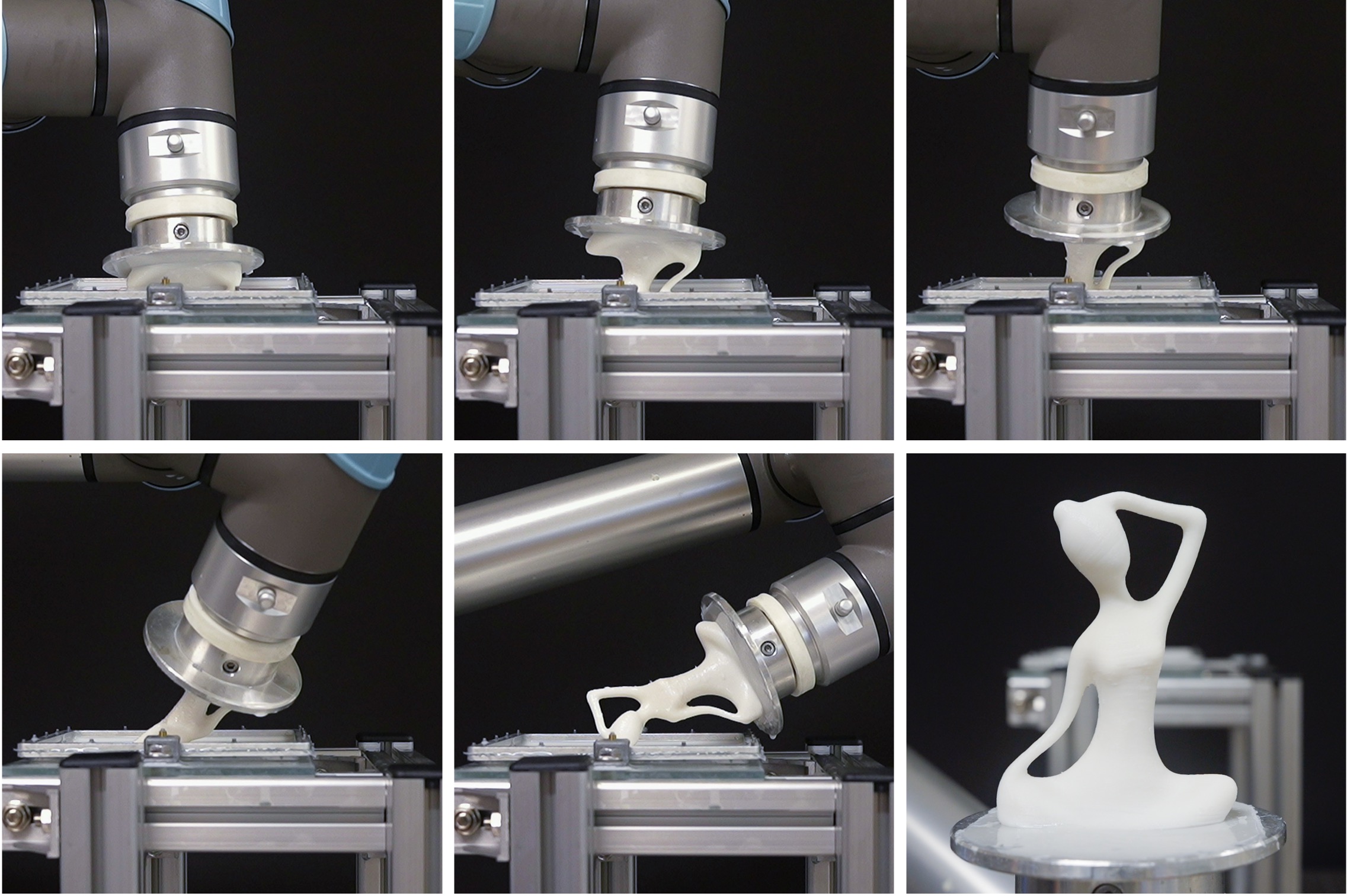}
\caption{The progressive results of printing the Yoga model by the robot-assisted multi-axis DLP hardware. 
}\label{fig:Yoga_progFab}
\end{figure}

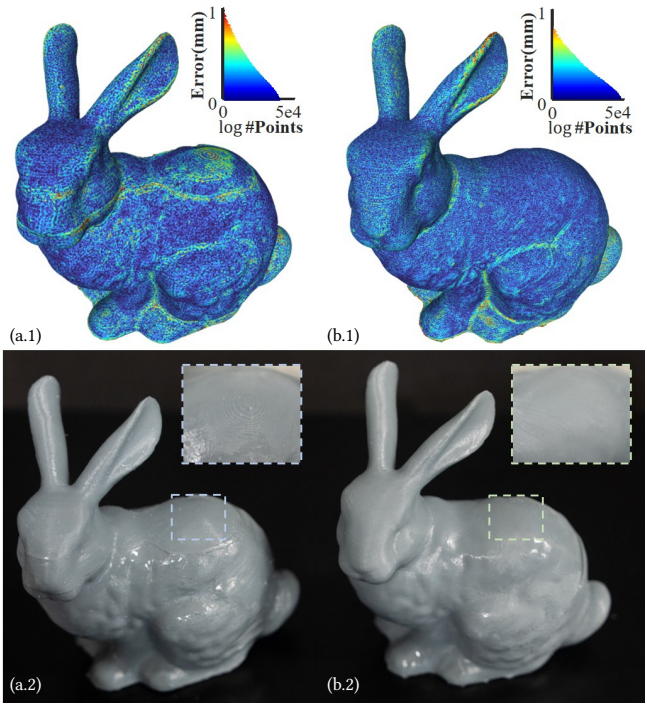
\begin{figure}[t]
\centering
\includegraphics[width=\linewidth]{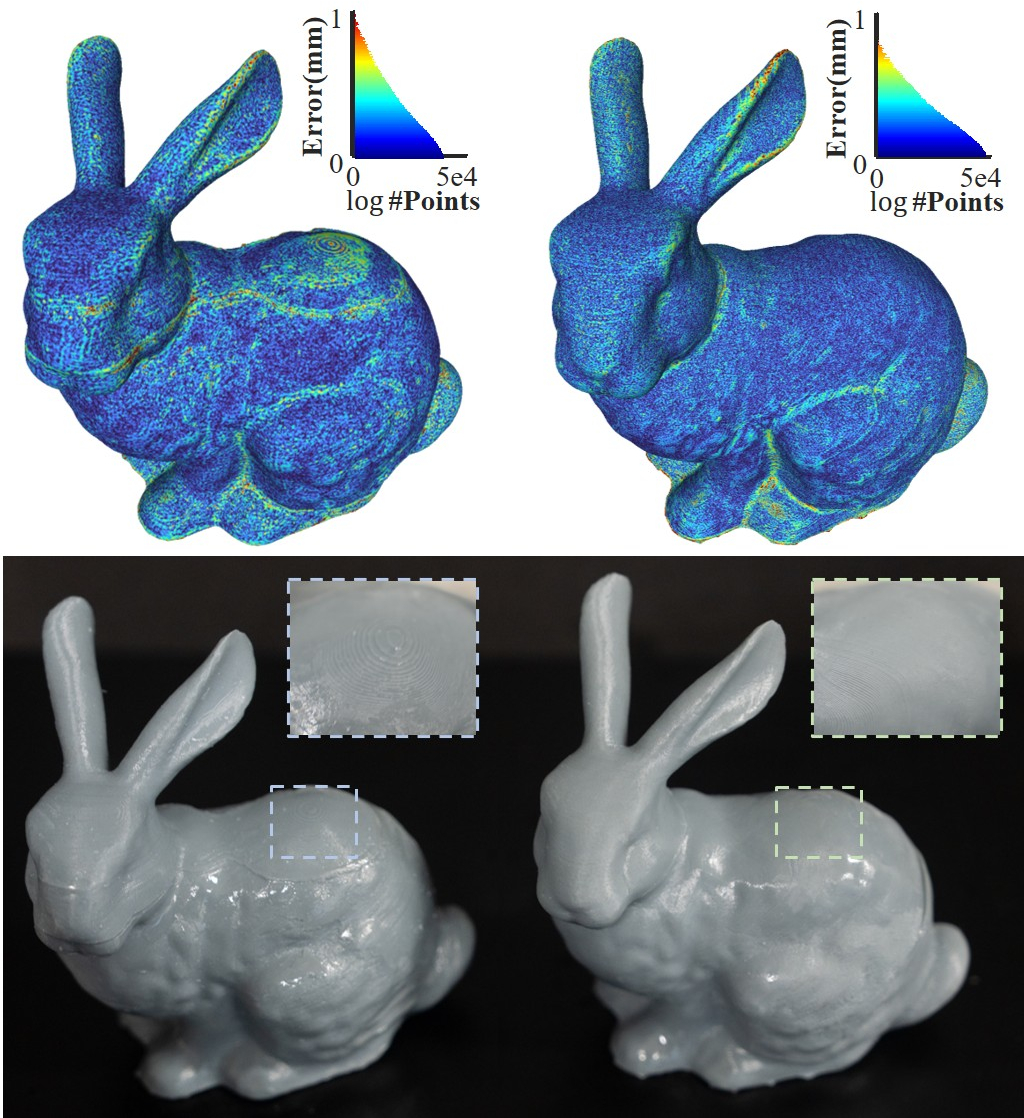}
\put(-240,137){\footnotesize \color{black}(a.1)}
\put(-120,137){\footnotesize \color{black}(b.1)}
\put(-240,5){\footnotesize \color{white}(a.2)}
\put(-120,5){\footnotesize \color{white}(b.2)}
\caption{By turning off the surface quality loss $\mathcal{L}^{SQ}$, the back of the Bunny model generate significant staircase artifacts (a.1 \& a.2) comparing to the model fabricated from our optimization using the full set of losses (b.1 \& b.2). The geometric errors on the scanned models are visualized as the color maps and the corresponding histograms in the top row (a.1 \& b.1).
}\label{fig:Bunny_phyAblation}
\end{figure}

\subsection{Physical Experiments}
The slicing results generated by our method were validated on a lab-built multi-axis DLP 3D printing setup, comprising a resin tank, a UV-light projector, and a UR5e robotic arm. The robotic arm is controlled via ROS2 and achieves a motion repeatability of $\pm 0.03\text{mm}$. The Sumaopai SMPDE75T4B projector emits UV light at a wavelength of $450\text{nm}$, covering an exposure area of $144\text{mm} \times 81\text{mm}$ as the effective area at the bottom of the resin tank (see Fig.~\ref{fig:collisionIllustration}). The curing process is digitally controlled using grayscale images at a resolution of $1920 \times 1080$. For physical fabrication, we used Anycubic Standard Resin V2 as the printing material in our experiments. 

\begin{figure}[t]
\centering
\includegraphics[width=\linewidth]{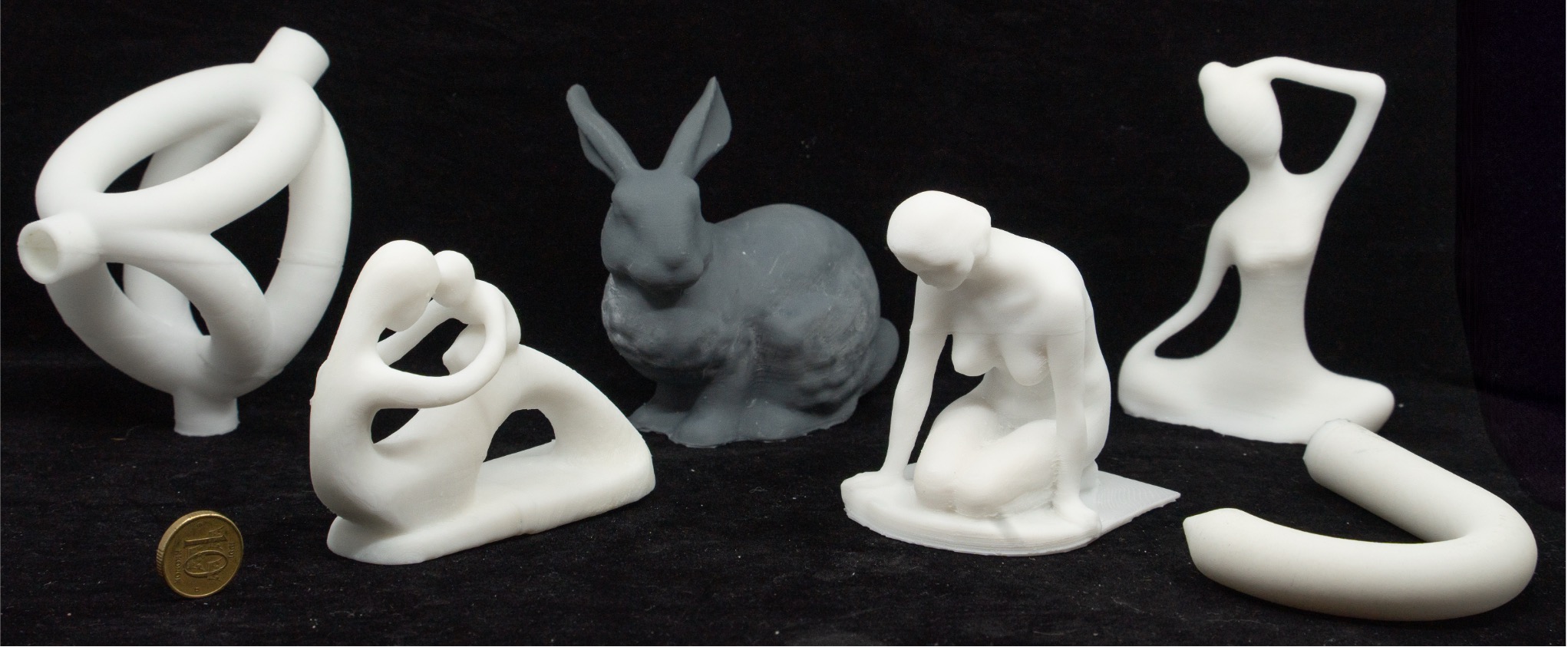}
\caption{The group photo of all models fabricated by our multi-axis DLP 3D printing method.
}\label{fig:progRes_GroupPhoto}
\end{figure}

All models discussed above were successfully fabricated using our multi-axis DLP 3D printing system. As illustrated in Figs.~\ref{fig:teaser} and \ref{fig:Yoga_progFab}, complex geometries can be effectively produced with our method. In addition to the computational ablation study presented in Sec.~\ref{subsubSec:AblationStudy}, we conducted a physical ablation study on the Bunny model to evaluate the impact of the surface quality loss term $\mathcal{L}^{SQ}$. The results are given in Fig.~\ref{fig:Bunny_phyAblation}, which demonstrate a significant reduction in staircase artifacts on the back of the Bunny model when $\mathcal{L}^{SQ}$ is applied. Final fabrication results for other models are shown in Fig.~\ref{fig:progRes_GroupPhoto}, and detailed fabrication statistics are provided in Table~\ref{tabCompStatistic}. In summary, all models were fabricated within 4 hours, with a total number of layers reaching up to 2.0k, demonstrating the efficiency of multi-axis DLP 3D printing.

\subsection{Discussion}
While our curve-based slicing algorithm successfully generates motion trajectories for multi-axis DLP 3D printing -- effectively overcoming manufacturing challenges such as collision, floating points, large overhangs, and staircase artifacts -- there remain several limitations to the current method.

\begin{figure}[t]
\centering
\includegraphics[width=\linewidth]{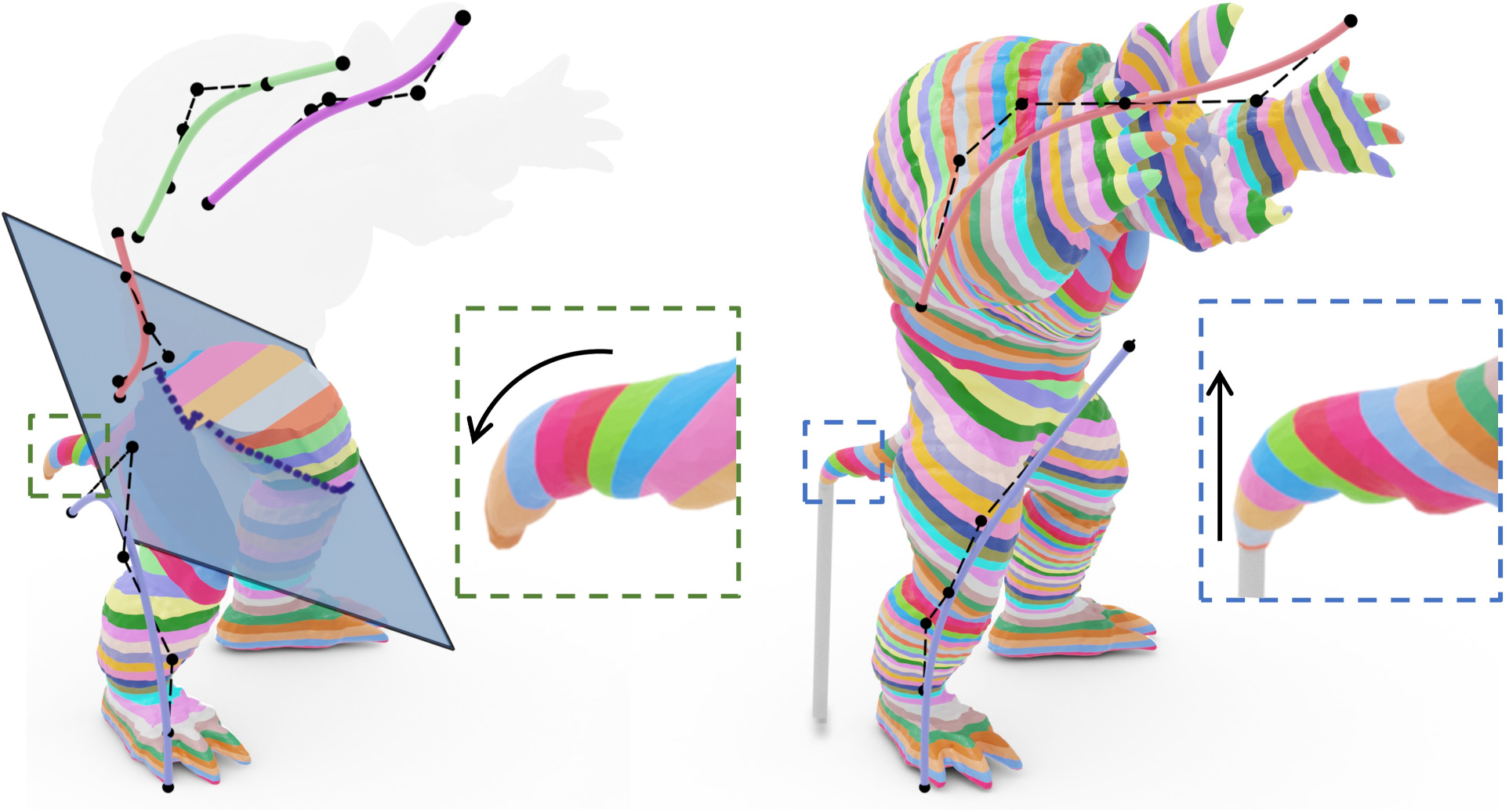}
\put(-240,5){\footnotesize \color{black}(a)}
\put(-115,5){\footnotesize \color{black}(b)}
\caption{A failure case caused by the collision between the tank and the already printed part when attempting to generate curves for support-free printing the tail of the Armadillo model (a). However, the model can become printable by simply adding a support to the tail (b).
}\label{fig:failureCase_Armadillo} 
\end{figure}

First, the manufacturability of the optimized curve trajectories can still be hindered by environmental obstacles, most notably the resin tank. For example, in the Armadillo model shown in Fig.~\ref{fig:failureCase_Armadillo}(a), the support-free trajectory for printing the tail results in a collision between the tank and the already-printed part, rendering the fabrication infeasible. However, this issue can be mitigated by introducing a single support under the tail (see Fig.~\ref{fig:failureCase_Armadillo}(b)), which enables successful printing while maintaining most of the benefits of support-free fabrication in other regions.

Second, our current collision-avoidance loss function (described in Sec.~\ref{sec:Collision}) only accounts for potential collisions between the printed structure (including the building plate) and environmental obstacles (e.g., the resin tank). It does not consider collisions involving the robotic arm and its surrounding environment. At present, we rely on the IK solver to discard unrealizable robot poses that would result in collisions. However, this approach does not guarantee the existence of feasible collision-free arm motions for all optimized trajectories.

\begin{figure}[t]
\centering
\includegraphics[width=\linewidth]{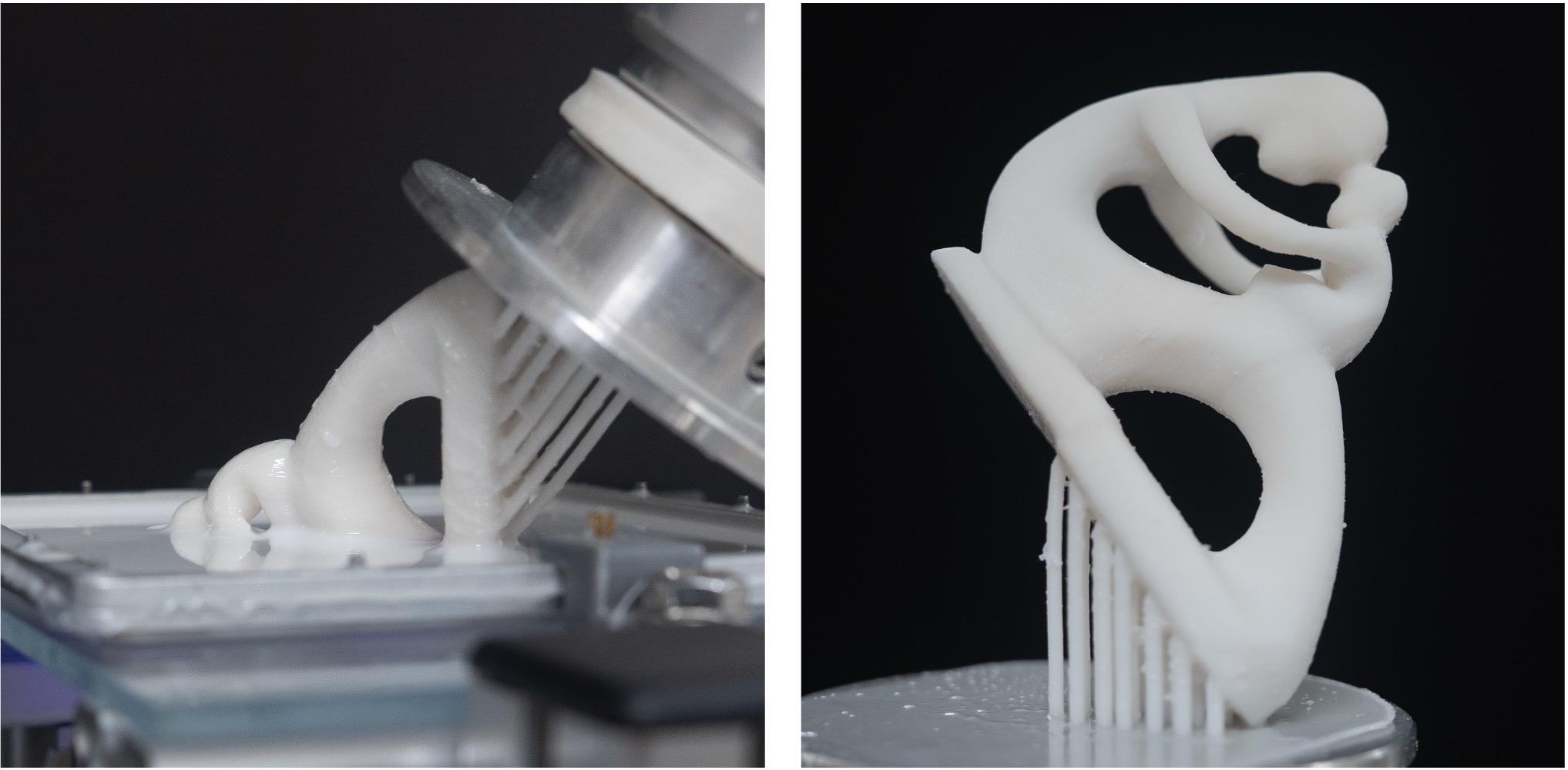}
\caption{Additional supports are manually added to the base of the Fertility model to improve the structural strength of the initial layers.
}\label{fig:fertility_support}
\end{figure}

Lastly, our method does not currently account for the mechanical stability of partially printed structures during the printing process (ref.~\cite{Zhou2013WorseCase}). For instance, additional supports were manually added to the base of the Fertility model to improve the structural integrity of the initial layers (see Fig.~\ref{fig:fertility_support}). The support structure is generated in two steps:
\begin{enumerate}
\item {A convex region is defined by the building plate’s plane and the first slicing plane at $\mathbf{c}(0)$;}

\item {Within this convex region, support material is added along the `vertical' direction, which corresponds to the normal vector of the build plate.}
\end{enumerate}
This method of support generation ensures collision-free DLP 3D printing during the initial stages of fabrication. Without these reinforcements, the connection between the early printed parts and the building plate would be too weak, resulting in unstable or failed prints. Furthermore, mechanical and thermal deformations occurring during the printing process may introduce misalignments between partitioned regions of the model. All of these limitations will be considered in our future work.

\section{Conclusion}
In this paper, we presented a novel curve-based slicer for jointly generating, 1) planar layers with dynamically varying orientations and 2) model partitions, to enable the fabrication of complex geometries using multi-axis DLP 3D printing. By introducing additional DoF in motion, our approach addresses several key limitations of conventional PPL-based DLP printing, including the need for support structures to handle floating points and large overhangs, as well as the presence of staircase artifacts on printed surfaces. 
Specifically, we define slicing layers using the tangent planes of one or more parametric curves, which also determine the model partitioning and the motion trajectories of the build platform through associated pose frames. Manufacturing objectives such as surface quality and overhang minimization, along with constraints like collision avoidance and floating-free deposition, are formulated as loss functions and jointly optimized through the control points of these curves. Additionally, the optimization of model's setup orientation can be integrated into the optimization framework. 
As a result, our method significantly enhances the capability of DLP 3D printing to handle complex geometries while preserving its intrinsic advantages of high resolution and fast printing speed. The effectiveness of our approach has been demonstrated through both numerical simulations and physical experiments.

\begin{acks}
    The authors gratefully acknowledge Professor A. John Hart and Nicholas S. Diaco of the Massachusetts Institute of Technology for their valuable comments on DLP printing. This research was supported by the InnoHK initiative of the Innovation and Technology Commission of the Hong Kong Special Administrative Region Government, the Chair Professorship Fund at the University of Manchester and UK Engineering and Physical Sciences Research Council (EPSRC) Fellowship Grant (Ref.\#: EP/X032213/1).
\end{acks}

\bibliographystyle{ACM-Reference-Format}
\bibliography{reference}
\end{document}